\magnification1200

\rightline{KCL-MTH-07-09}

\vskip 2cm \centerline {\bf  Duality Symmetries  and $G^{+++}$  Theories } \vskip 1cm \centerline{Fabio
Riccioni, Duncan Steele and Peter West} \centerline{Department of Mathematics}  \centerline{King's College,
London WC2R 2LS, UK}

\vskip 2.5cm
\vskip .2cm \noindent We show that the non-linear realisations of all the   very extended algebras $G^{+++}$,
except the $B$ and $C$ series which we do not consider,  contain fields corresponding to all possible duality
symmetries of the on-shell degrees of freedom of these theories. This result  also holds for $G_2^{+++}$ and
we argue that the non-linear realisation of this algebra accounts precisely for the form fields present in
the corresponding supersymmetric theory. We also find a simple necessary condition for the roots to belong to
a $G^{+++}$ algebra. \vskip .5cm

\vfill \eject

\noindent The IIA [1] and IIB [2,3,4] supergravity theories, by  virtue of the large amount of supersymmetry
that they possess, encode all the low energy  effects of their  corresponding string theories, while the
eleven dimensional supergravity theory [5]  is thought to be the low energy  effective  action for an as yet
undefined theory called M theory. Given that there  exists no complete  formulation of string theory these
supergravity theories have proved invaluable in our understanding of string theory  and extensions of it to
include branes. The result [6] that these supergravity theories can be formulated as  non-linear realisations
led to the conjecture [7] that the non-linear realisation of the generalised Kac-Moody algebra $E_{11}$ was
an extension of eleven dimensional supergravity. Indeed this algebra, when decomposed  with respect to its
$A_{10}$, or $SL(11)$, algebra associated with eleven dimensional gravity, has lowest level  positive root
generators which are in one to one correspondence with the graviton, the three and six form gauge  fields,
while the next generator corresponds to the dual graviton [7].  Non-linear  realisations of $E_ {11}$ can
also be used to describe the maximal ten dimensional supergravity theories, but in this case  the  adjoint
representation is decomposed into representations of $A_{9}$, or $SL(10)$, associated with ten  dimensional
gravity. There are only two such $A_9$ subalgebras and the two different choices were found to  lead to non-
linear realisations which at low levels are the IIA and IIB supergravity theories in references [7] and [8]
respectively. It is striking to examine tables of the generators [9] listed in terms of increasing level  and
see  how the generators associated with the field content of the IIA and IIB supergravity theories occupy
precisely  all  the lower levels, before an infinite sea of generators whose physical significance was
unknown at the time the tables of reference [9] were constructed. The fact that the three maximal
supergravity theories in  ten and eleven dimensions could be formulated in terms of a single $E_{11}$ theory
encouraged the belief [7] that this algebra might be a symmetry of the underlying M theory.

Amongst the set of $E_{11}$ generators appropriate to  the IIA theory  is one with nine antisymmetrised
indices which in the non-linear  realisation leads to a field with the same  index structure [9]. This nine
form generator occurs in the table of IIA generators at a place which is  in amongst the generators
associated with the fields  of  the  IIA supergravity theory. A non-trivial value for  this  field is known
to lead to the massive IIA supergravity theory [10] and as a result it was realised  [9] that $E_{11}$ can
incorporate the massive IIA  theory.

However, there is a one to one correspondence between the fields of  the non-linear realisations of $E_{11}$
appropriate to the eleven dimensional, IIA and IIB theories and using  this correspondence one finds  that
the nine form of the IIA  theory corresponds to a field with the index  structure $A^{a,b,c_1\ldots c_{10}}$
which occurs at a level which is beyond those of the supergravity fields of  eleven dimensional supergravity
theory [11]. As such $E_{11}$ provides an eleven dimensional origin of the  massive IIA theory which involves
one of the higher level fields and in this way the physical interpretation of at  least one of the higher
level fields in the eleven dimensional $E_{11}$ theory became apparent.

For quite some time the significance of  the higher fields in the non- linear realisation,  with the
exception of the one field just mentioned above,  was unclear.  However, it was  shown that the quadruplet
and doublet space-filling forms of the IIB theory which were known [9] to be  present in the $E_{11}$
formulation were found from a rather different perspective. Remarkably it was shown [12]  that if one
included the dual fields corresponding to  all the physical fields of the IIB supergravity  theory then the
supersymmetry algebra could be closed in the presence of a set of ten forms which were  precisely  the fields
that were predicted by $E_{11}$. Furthermore, it was shown that there was a precise matching  of coefficients
for the gauge algebras of all these forms found on the one hand from closing the supersymmetry  algebra and
on the other hand from the $E_{11}$ algebra [13]. A similar story applies to the closure of the  IIA
supergravity algebra [14] and the space filling forms predicted from $E_{11}$ [9].

Recently considerable numbers of higher level $E_{11}$ fields were  shown to have a physics meaning. First it
was shown [15] that an infinite  class of the higher fields were just  those required to realise all possible
duality symmetries of the basic on-shell degrees of freedom of the theory.  More recently all the form
fields, that is those fields whose indices are totally antisymmetrised, resulting  from the dimensional
reduction of the eleven dimensional non-linear realisation were calculated [16]. It emerged  that the
formulation of the maximal supergravities in the lower dimension $D$ which arose from $E_{11}$  was
democratic meaning that the physical degrees of freedom of the theory were described by two fields whose
field strengths were related by Hodge duality, except in the case of self-dual fields. Furthermore, the  rank
$D-1$ forms that arose could  be used to classify all possible deformations of the maximal supergravity
theories that arise from a Lagrangian formulation and we found that the results were in complete agreement
with the  known maximal gauged supergravity theories. The latter have been found over many years by
considering the  deformations of the massless maximal supergravity theory in  the dimension of interest (see
for example reference [17]  and references therein). This result also provided an  eleven dimensional origin
for all the  gauged, or  massive, maximal supergravity theories. Reference [16] also found all the
space-filling forms which are important for  the consistency of orientifold models [18]. The results of
reference [16] were also found in reference [19] by  analysing the $E_{11}$ theory directly in the dimension
of interest.

Arguments similar to those advocated  for eleven dimensional   supergravity in [7] were applied to  the
effective action of the closed bosonic string  D dimensions [7],  to gravity in  D dimensions [20], and the
type I supergravity theory [21] and the underlying Kac-Moody algebras were  identified. It was realised that
the algebras that arose in all these theories were of a special kind and  were called very extended Kac-Moody
algebras [22]. Indeed, for any finite dimensional semi-simple Lie  algebra $\cal G$ one can systematically
extend its Dynkin diagram by adding three more nodes to obtain an indefinite  Kac-Moody algebra denoted $\cal
G^{+++}$. In this notation $E_{11}$ is written as $E_8^{+++}$. The Kac-Moody  algebras that were conjectured
to underlie the closed bosonic string, gravity and type I supergravity being $D^{+ ++}_{D-2}$ [7],
$A^{+++}_{D-3}$ [20], and $D^{+++}_8$ [21] respectively.

The  ideas in  [7,20,21,22] were generalised in references [23,24] to  consider the non-linear realisation of
any ${\cal G}^{+++}$ algebra. For each very extended algebra ${\cal G}^{++ +}$ one can find an $A_{D-1}$
sub-algebra that is associated with a set of $D-1$ nodes  which form a line in  the ${\cal G}^{+++}$ Dynkin
diagram. This line is called the gravity line and it must start with the very  extended node. In the
non-linear realisation the $A_{D-1}$ sub-algebra is associated with the gravity sector of the  theory and the
resulting theory lives in $D$ dimensions. In general there is more than one possible gravity line,  or
$A_{D-1}$ sub-algebra. By analysing ${\cal G}^{+++}$  with respect to  the $A_{D-1}$ sub-algebra it was
shown [9] that the low level fields present in the non-linear realisation of  ${\cal G}^{+++}$ contained
gravity,  form gauge fields and in some cases scalars which was the result anticipated by the above
conjectures.  This was also consistent with the oxidation points [24] of all the three dimensional theories
which possessed  coset symmetries.

In this paper we investigate if the $\cal G^{+++}$ algebras have some  of the same features described above
for the $E_8^{+++}$ algebra. In particular we will see that if the  gravity sub-algebra of ${\cal G}^{+++}$
being considered is  $A_{D-1}$ then we can define dual generators to be  those that  have no blocks of
anti-symmetrised $D$ or $D-1$ indices. We will then find all dual generators and  show  that they are only
those corresponding to the on-shell degrees of freedom of the theory together with those  fields whose field
strengths are related to these by Hodge duality, as well as generators that have the same  index structure as
these fields, but are decorated by any number of blocks of $D-2$ indices. We will show this  result for all
${\cal G}^{+++}$ algebras with the exception of the $B$ and $C$ algebras which we do not  consider. This
result implies that the non-linear realisation encodes all possible duality symmetries of the on-shell
degrees of freedom of these theories.

We also derive the form fields of the ${\cal G}^{+++}_2$ theory in  five,  four and three dimensions and
discuss the possible resulting deformations of the theory. The results are  completely compatible with the
know literature on the subject. In two separate appendices, we also derive the form fields arising in any
dimension from 8 to 4 for the $E_6^{+++}$ theory and from 6 to 4 for the $F_4^{+++}$ theory, while a third
appendix is devoted to a discussion of the $G_2^{+++}$ case.

\vskip 1cm \noindent {\bf {1 \ \ $ E_{8}^{+++}$}}
\medskip
\noindent Any generator in a Kac-Moody algebra can by definition be  written as the multiple commutators of
the Chevalley generators ( $H_a , E_a$ and $ F_a$) which are subject to  the Serre relations. However for no
general Kac-Moody algebra is even a complete listing of the generators known.  A Lorentzian Kac-Moody algebra
[22] is one where the deletion of at least one node in the Dynkin Diagram will  leave a finite dimensional
algebra and at most one affine algebra. These algebras are more tractable, in that  their properties can be
analysed in terms of the well understood algebra that remains after the deletion of the  preferred node in
the Dynkin diagram.  In particular, one can analyse the adjoint representation of the Kac- Moody algebra in
terms of the representations of the algebra that remains. The level of a given generator of the  Kac-Moody
algebra is defined to be the number of times the generator corresponding to the deleted node occurs in  the
multiple commutator in which the generator arises. Often one can delete a node such that the algebra  that
remains is $A_n$  and if this is not the case one can delete another node to find an $A_n$  algebra. In  this
section we will review how the determination of the properties of the Kac-Moody algebra are  determined in
terms of the $A_n$ sub-algebra [22,25,26] and recall how it works in detail for the case of $E_{11}$  [26].
We will then recover the dual generators of reference [15], but also give some new insights into   how to
determine the remaining generators of the algebra.

For an algebra where only one node, labelled by $c$, needs to be  deleted to find an $A_n$ gravity
subalgebra, the simple roots of the Kac-Moody algebra can be written in terms of  those of $A_n$, that is
$\alpha_i, \ i=1,\ldots , n$, and the simple   root corresponding to the  deleted  node which can be written
as  [22]
$$ \alpha_c = x - \nu \eqno(1.1)$$
where  $\nu$ is given by
$$
\nu= -\sum_i A_{ci}\lambda _i  \ , \eqno(1.2)$$ $x$ is a  vector  orthogonal to the $A_n$ weight lattice and
the $\lambda_i$ are the fundamental weights of the $A_n$ subalgebra.  Indices $a,b,\ldots $ run over the rank
of the full Kac-Moody algebra, while   $i,j,\ldots $ over the rank  of the  $A_n$ sub-algebra. We note that
the simple roots do indeed replicate the Cartan matrix $A_{ab}$  of the Kac-Moody algebra, which if this is
symmetric are given by $(\alpha_a, \alpha_b) = A_{ab}$. Here we used that $ (\alpha_i, \lambda_j) =
\delta_{ij}$ for the $A_n$ sub-algebra as indeed is the case for all simply laced finite  dimensional
semi-simple Lie algebras.  The quantity $x^2$ is determined by demanding that $\alpha_c^2 $ have the  correct
value, which for the case of symmetric Cartan matrix is just $\alpha_c^2 = 2$.

Any root of the  Kac-Moody  algebra can be written, using equation  (1.2), in the form
$$
\alpha =\sum _i n_i\alpha_i + l\alpha_c = lx - \Lambda,\ \  {\rm   where }\ \ \Lambda  = \nu -  \sum_i
n_i\alpha_i . \eqno(1.3)$$ We recognise $l$ as the level.  We see  that $\Lambda$ belongs to the weight
lattice of the $A_n$ subalgebra. If a representation of $A_n$, with highest  weight $\sum _i p_i\lambda_i$,
where $p_j$ are the Dynkin indices, occurs then this  highest weight must occur  as one of the possible
$\Lambda$'s as the roots of the Kac-Moody algebra vary. As such, a necessary condition  for the adjoint
representation of the generalised Kac-Moody algebra to contain the highest weight  representation of $A_{n}$
with Dynkin indices $p_j$ [22, 25,26] is that
$$\sum _j p_j \lambda_j = l\nu -\sum _i n_i\alpha_i \ .
\eqno(1.4)$$ Taking the scalar product with $\lambda_k$ leads to  the  condition
$$ n_k = l\nu\cdot\lambda_k - \sum _j p_j(\lambda_j,
\lambda_k) \ . \eqno(1.5)$$ We recall that  $(\lambda_i, \lambda_j) =  A_{ij}^{-1}$ for any  simply laced
algebra, where $A_{ij}^{-1}$ is the inverse Cartan matrix which is  positive definite for a finite
dimensional semi-simple Lie algebra.  The inverse Cartan matrix of the $A_{D-1}$  algebra is given by
$$ A^{-1}_{ij} = \cases{
{i(D-j) \over D}, &$i\leq j$\cr {j(D-i) \over D}, &$j\leq i$\cr }  \eqno(1.6)$$ In equation (1.5) the
integers $l,n_k,p_j$ must be positive  and so this places a   necessary, but  not sufficient, condition on
the possible $A_n$ representations   contained in the adjoint representation of  the Kac-Moody algebra at
level $l$.

Taking the scalar product of the expression for $\alpha$ of equation  (1.3)  we find that
$$ \alpha^2 = l^2x^2 + \sum _{i,j} p_i (\lambda_i,\lambda_j) p_j =2, 0,
-2, \eqno(1.7)$$ We have used the fact that the  lengths of the roots  of  a symmetric Kac-Moody    algebra
are constrained to  take the values $ 2, 0, -2,...$ [27].

Thus we find  a second necessary, but not sufficient, constraint on  the $A_n$ representations contained in
the adjoint representation of the Kac-Moody algebra.  In fact the    two  constraints of equations (1.5) and
(1.7) are not as strong as imposing the Serre relations on the multiple  commutators, although almost all the
solutions they possess are  roots that actually appear in the Kac-Moody algebra.

Let us explain this procedure for the case of  $E_{11}$ [26] whose  Dynkin diagram is given below
$$
\matrix{ & & & & & & & & & & & & & &0&11&\cr
        & & & & & & & & & & & & & &|& &\cr
       0&-&0&-&0&-&0&-&0&-&0&-&0&-&0&-&0&-&0\cr
       1& &2& &3& &4& &5& &6& &7& &8& &9& &10\cr}
$$

Deleting  the  node 11  leaves an $A_{10}$ subalgebra.  The simple  roots of $E_{11}$ are those of $A_{10}$,
i.e. $\alpha_i,\ i=1,\ldots 10$,  as well as the simple root of the  deleted node which is given by
$$
\alpha_{11}=x-\lambda_8,\ \ \ x^2=-{2\over 11} \eqno(1.8)$$ The  general root of $E_{11}$ has the form
$$\alpha=l\alpha_{11}+\sum_{i=1}^{10} n_i \alpha_i
=lx-\Lambda \eqno(1.9)
$$
where $\Lambda= l\lambda_8-\sum_{i=1}^{10} n_i \alpha_i$. This  equation effectively rewrites the adjoint
representation of $E_{11}$ in terms of representations of $A_{10}$. A  necessary condition for a
representation of $A_{10}$ to occur is that the set of all $\Lambda$'s contain the  highest weight of the
representation in question, that is $\sum_i p_i \lambda_i$, where the   $p_i$ are the  Dynkin  indices. As
such,  we can set $\Lambda=\sum_i p_i \lambda_i$ and taking the  scalar product with $ \lambda_j$ we find
that
$$
\sum_i p_i \lambda_i\cdot \lambda_j-l  \lambda_8\cdot \lambda_j=-n_j  \eqno(1.10)$$ While the square of the
corresponding $E_{11}$ root is given by equation (1.7), that is
$$ \alpha^2 = -{2\over 11} x^2 + \sum _{i,j} p_i (\lambda_i,  \lambda_j) p_j
=2, 0, -2,... \eqno(1.11)$$

Any  generator is found by taking the multiple commutator of the  Chevalley generators and the level of a
generator is the number of times the Chevalley generator   corresponding to node 11 appears in this multiple
commutator. The Chevalley  generator associated with node 11  has  three $A_{10}$ indices, hence it adds
three indices every time it appears in the multiple commutator. Therefore  any level $l$ generator can be
written with $3l$ indices. We note that the  remaining Chevalley generators are  contained in the $K^a{}_b$
generators of $A_ {10}$ and so these do not change the number of indices.  Given a  generators with Dynkin
index $p_i$, the index contributes $p_j$ blocks of  $11-j$ anti-symmetrised indices; as a  result the total
number of indices obeys the relation [15]
$$
3l=\sum_i(11-i)p_i +11 \ m \eqno(1.12)$$ The last term corresponds to  the possible presence of $m$ blocks of
fully antisymmetrised rank 11 indices. Such  blocks do not transform  under the $A_{10}$ subalgebra.

If we substitute the level condition of equation (1.12) into the root  length condition of equation (1.11) we
find that
$$
\alpha^2= {1\over 9} \sum _{j=1}^{10} (11-j)(j-2) p_j^2 +{2\over 9} \sum_{i<j}(11-j)(i-2) p_i p_j -{4\over 9}
m \sum _i (11-i)p_i -{2.11\over 9} m^2
$$
$$=2,0,-2,\ldots
\eqno(1.13)$$ Substituting the level condition of equation (1.12)  into the root condition of equation (1.10)
we find that the factors of $1\over 11$ disappear and the values of $n_i $ are given by
$$
n_j=\sum _{i, i<j} (j-i) p_i +jm, \ j=1,\ldots ,8, \ \ n_9= {2\over 3} (\sum_i (8-i)p_i+8m) + p_{10},\ \
$$
$$
n_{10}= {1\over 3} (\sum_i (8-i)p_i+8m ) \eqno(1.14)$$ We see that  the $n_i,\ i=1,\ldots , 9$ are positive
as they must be. Furthermore, using the level condition of equation  (1.12) one sees that $\sum_j
(8-j)p_j+8m$ is a multiple of $3$ as $3(l-\sum_jp_j -s)=\sum_j (8-j)p_j+8m$ and so $n_ {9}$ and $n_{10}$ are
integers which one can also show are  positive.

Hence, we find the perhaps surprising result  that the root condition  of equation (1.10) is automatically
satisfied if we use the level matching condition of equation  (1.12).  We recall that previously one found
the possible  roots of $E_{11}$ by finding all  the solution of equations (1.10) and (1.11). However, now we
need only solve  the level matching condition of  equation (1.12) and the  reformulation of the length
squared condition of equation (1.13). Clearly, these are much simpler conditions that those of equations
(1.10) and (1.11).

In reference [15] the concept of dual generators was introduced;  these are generators which possess  with no
blocks of rank 10 or 11 totally antisymmetrised indices.  This means  that $m=0=p_1$ and so now  $3l= \sum_j
p_j (11 -j)$. We observe that in the root length condition of equation  (1.13) all terms on the right-hand
side become positive and that as $p_2$ is absent this Dynkin index  has no  condition placed on it. Taking
$p_2=0$, the only allowed roots  are given by
$$\eqalign{
\alpha_A &= (0,0,0,0,0,0,0,0,0,0,1),\  p_8 = 1\cr \alpha_B &=  (0,0,0,0,0,1,2,3,2,1,2),\ p_5 = 1\cr \alpha_C
&= (0,0,0,1,2,3,4,5,3,1,3),\  p_3 = p_{10} = 1\cr } \eqno(1.15)$$ The  other $p_i$s are 0. These correspond
in the non-linear  realisation to the three form, six form and dual graviton  generators respectively  and as
expected they occur with multiplicity one. The solution for $p_2=1$ is given by
$$
\gamma = (0,0,1,2,3,4,5,6,4,2,3),\  p_2 = 1 \eqno(1.16)$$ This  solution has multiplicity zero  and so does
not actually occur in the $E_{11}$ algebra. It corresponds to a generator  of the form $R^{a_1\ldots a_9}$.

The roots corresponding to all the dual generators can then be    written in the form [15]
$$\eqalign{
\alpha_A(p_2) =  \alpha_A+p_2\gamma\cr \alpha_B(p_2) =   \alpha_B+p_2 \gamma\cr \alpha_C(p_2) =
\alpha_C+p_2\gamma\cr \alpha_D(p_2) = p_2 \gamma\cr } \eqno(1.17)$$ They  are just  found by taking multiple
commutators of the multiplicity generator $R^{a_1\ldots a_9}$ with each of the  generators corresponding to
the roots of equation (1.15).

The new fields found in the non-linear realisation corresponding to the roots of equation (1.17) differ from
those of equation (1.15) by blocks of nine indices. However, as the  little group of the massless states is
SO(9) these fields  describe the same on-shell states.   As a result, we find that the $E_{11}$ non-linear
realisation includes all possible ways of describing the original degree of  freedom of the theory and so we
may conclude that $E_{11}$ encodes all possible duality transformations  [15].

We now apply the above arguments to the other very extended algebras.

\vskip 1cm \noindent {\bf {2 \ \ $ A_{D-3}^{+++}$}}
\medskip
\noindent It has been conjectured that a  suitably extended version  of pure gravity in $D$ dimensions can be
expressed as a nonlinear realisation of $A_{D-3}^{+++}$ [20].   Analysing this algebra with respect to its
$A_{D-1}$ sub-algebra we find at level 0 the $A_{D-1}$ generators $K^a {}_b$, corresponding to the graviton,
and at level 1 the generator $R^{a_1 \dots a_{D-3},b}$ corresponding to  the dual graviton field. The Dynkin
diagram of $A_{D-3}^{+++}$ is given by

$$
\matrix{ & & & & & & &D& & & \cr
       & & & & &-&-&0&-&-& \cr
       & & & &|& & & & & &|&\cr
       & & & &|& & & & & &|& \cr
      0&-&0&-&0&-&0&\cdots&0&-&0&\cr
      1& &2& &3& & & & & &D&-1&\cr}
$$
For simplicity we will assume that $D>4$. Deleting the node labelled $D$ we find the $A_{D-1}$ gravity
subalgebra. The simple roots of $A_{D-3}^{+++}$ can be written as the  simple roots $\alpha_i,\ i=1, \ldots
D-1$ of $A_{D-1}$ as well as the simple root of  the deleted node which  is given by
$$\alpha_D = x - \lambda_3 - \lambda_{D-1} \ .
\eqno(2.1)$$ Given $\alpha^2 = 2$ we find $x^2 = {4 \over D} - 2$. As  such,  a  general root of
$A_{D-3}^{+++}$ can be written as
$$
\alpha = \sum_i n_i\alpha_i + l\alpha_D = lx - \Lambda ,\ \ {\rm  where}\ \ \Lambda = l(\lambda_3 +
\lambda_{D-1}) - \sum_i n_i\alpha_i \ . \eqno(2.2) $$

A highest weight representation of  $A_{D-1}$ with Dynkin indices $p_i $ can occur if $\Lambda = \sum_i p_i
\lambda_i$ occurs. Taking the scalar product of this equation with $ \lambda_k$ gives an equation for the
root components
$$ n_k = l\left((\lambda_3,\lambda_k) + (\lambda_k,\lambda_{D-1})  \right) -
\sum_jp_j(\lambda_j,\lambda_k) \ .  \eqno(2.3)$$

Squaring the expression for $\alpha$ in (2.2) gives
$$
\alpha^2 = -{2(D-2) \over D} l^2 + \sum_{i,j}  p_i (\lambda_i, \lambda_j) p_j = 2,0,-2,\dots \eqno(2.4)$$

The Chevalley generator corresponding to the  deleted node $D$ has a  block of $D-3$ antisymmetrised
$A_{D-1}$ indices, and a vector index, i.e. $R^{a_1 \dots a_{D-3}, b}$, giving  $D-2$ indices overall. A
level $l$ generator, the multiple commutator of which contains the generator  corresponding to the deleted
node $l$ times, will therefore  have $ (D-2)l$ $A_{D-1}$ indices in total.   Hence we  get
$$ \sum_j p_j (D-j) + Dm = (D-2)l
\eqno(2.5) $$ where $m$ is the number of blocks  of $D$  antisymmetrised indices. Substituting this
expression for $l$ into (2.4) we find that
$$ \alpha^2 =  {1 \over (D-2)} \left( \sum_j p_j^2 (D-j)(j-2) + 2
\sum_{i<j} p_ip_j(D-j)(i-2) \right)
$$
$$
-{4m\over (D-2)} \sum_i(D-i)p_i -{2m^2 D\over (D-2)} = 2,0,-2,\dots \eqno(2.6)$$

We now show that any generator that satisfies (2.5) automatically  satisfies the root condition (2.3), namely
that it gives positive  integer values for the $n_k$.  Substituting  (2.5) into (2.3) we find that the
factors of ${1\over D}$ disappear and it gives
$$
n_j=\cases{p_{j-1}+jm,\ \ \ \ \ j=1,2,3\cr \sum_{i,i<j} (j-i)p_i +jm - (j-3)l,\ \ \ \ j=3,\ldots D-1 \cr}
\eqno(2.7)
$$
where we have formally taken $p_0=0$.

The definition of a dual generator  is that it has no blocks $D$ or  $D-1$  totally antisymmetric $A_{D-1}$
indices. This may be written $p_1 = m = 0$ and so the level matching  condition becomes
$$ \sum_j p_j (D-j)  = (D-2)l \ ,
\eqno(2.8) $$ while equation (2.6) becomes
$$ \alpha^2 =  {1 \over (D-2)} \left( \sum_j p_j^2 (D-j)(j-2) + 2
\sum_{i<j} p_ip_j(D-j)(i-2) \right) = 2,0,-2,\dots \eqno(2.9)$$

Two things are immediately obvious from this equation; $p_2$ is unconstrained, and both terms are positive
definite, so $\alpha^2$ can only be 2 or 0. For $\alpha^2 = 0$, both  terms must be 0, so only $p_2$ may be
nonzero.  Equation  (2.5) then implies that $l = p_2$ and equation  (2.7) gives the first $D-1$ coefficients
($n_1 \dots n_{D-1}$) of the root, while the last coefficient of the  root is simply the level, $n_D = l =
p_2$. Defining $\gamma$ to correspond to the coefficients $n_j=(0,0,1,\dots, 1,\dots,1)$, gives the roots
corresponding to all $\alpha^2 = 0$ dual generators in terms of the free parameter  $p_2$ by
$$
\alpha_B(p_2) = p_2 \gamma \ . \eqno(2.10)$$ In fact the generator with $p_2=1$ has multiplicity zero and so
does not appear in the algebra.

The remaining solutions have  $\alpha^2 = 2$.  We notice that $p_3 =  p_{D-1} = 1$, with $p_2=0$  and all
other $p_i=0$ gives a solution with $\alpha^2 = 2$. Using equation  (2.7)  we find that the corresponding
root is given by
$$
\alpha_A =   (0,0,...,0,1),\ \ p_3=1=p_{D-1} \ . \eqno(2.11)$$ This  corresponds to the generator
$R^{a_1\ldots a_{D-3},b}$ which has multiplicity one and so up to this level the   theory contains gravity
and the dual graviton [20,9].

Since $p_2$ does not appear in  equation (2.6) given any solution we  may find a whole class of solutions
that have the same Dynkin indices as before,  but with the addition of a $p_2$ that can be any positive
integer. Adding such a  $p_2$ leads to a root also satisfies the level  matching condition of equation (2.8)
as $p_2\to p_2+1$ just change $l\to l+1$. We recall that  equation (2.3) is  automatically satisfied if the
level matching condition holds.  Applying this to the solution of equation (2.11) we  have new solutions
which  can be written in the form
$$
\alpha_A(p_2) =\alpha_A  +p_2 \gamma \eqno(2.12)$$ This line of  argument will apply to all the $G^{+++}$
algebras  considered in this paper.

In fact all possible dual generators are given in equations (2.10)  and (2.12)  and summarising we find that
all the  dual generators present in the $A_{D-3}^{+++}$ algebras may  be written as
$$\eqalign{
\alpha_A(p_2) &= \alpha_A +p_2\gamma \cr \alpha_B(p_2) &=
 p_2\gamma  \cr } \eqno(2.13)$$

They encode all possible ways of describing the on-shell degrees of freedom of the theory which are just
those of gravity. As  such the dual fields encode all possible duality  transformations.

We now explain why there are no other solutions. The contribution from a single $p_i$ to $\alpha^2$ is given
by
$$
\alpha^2=p_i^2 \left((i-2) -{(i-2)^2\over (D-2)}\right) \ . \eqno (2.14)$$ Let us suppose that a single $p_i$
is non-zero and takes the value $p_i=r$. The  level matching condition  of equation (2.8) becomes
$(D-2)l=(D-i)r$ which in turn implies that $(D-2)(l-r)=-(i-2)r$. Using this latter  condition and that
$\alpha^2=2$ we find that $2=rl(i-2)$ which has no solution that is compatible with the level  matching
condition.

Now let us suppose that two Dynkin indices are non-zero, i.e. $p_i=1=p_j$ with $i<j$. In this case the
contribution to $\alpha^2$ is given by
$$
\alpha^2=3(i-2)+(j-2)-{(i-j)^2\over (D-2)} \ . \eqno(2.15)$$ However,  since $j-i<D-3$ we find that this
implies that $\alpha^2>2i-8$ and since $\alpha^2=2$ we conclude that $i=3$ is the only allowed value. Thus we
only find the one solution of equation (2.11).

\vskip 1cm \noindent {\bf {3 \ \  $E_7^{+++}$}}
\medskip
\noindent The non-linearly realised ${E_7^{+++}}$ theory contains at  low levels either a ten dimensional
truncation of IIB supergravity or an eight dimensional truncation of  maximal $D=8$ supergravity, depending
on the choice of gravity line [9]. Here we will only consider the ten  dimensional theory. The Dynkin diagram
of ${E_7^{+++}}$ is given by

$$
\matrix{ & & & & & & & & & & &0&10&     \cr
       & & & & & & & & & & &|&   \cr
       &0&-&0&-&0&-&0&-&0&-&0&-&0&-&0&-&0 \cr
       &1& &2& &3& &4& &5& &6& &7& &8& &9 \cr
} $$ The node labelled three is the affine node  with nodes one and  two being the over and very extended
nodes respectively.

In the case of $E_7^{+++}$ node ten is deleted to leave an $A_9$  subalgebra. The simple root corresponding
to node ten can be written as
$$ \alpha_{10} = x - \lambda_6 \eqno(3.1)$$
where $\lambda_6$ is a fundamental weight of the $A_9$ subalgebra.  Given that $\alpha_{10}^2 = 2$ we find
$x^2 = -2/5$. A general root of $E_{7}^{+++}$ can be written in terms of the  simple roots
$$
\alpha = \sum_{i=1}^9 n_i\alpha_i + l\alpha_{10} = lx - \Lambda, \ \   {\rm where}\ \ \Lambda = l\lambda_6 -
\sum_i n_i\alpha_i \ . \eqno(3.2)$$ A highest weight representation  of $A_{D-1}$  with Dynkin indices $p_j$
can occur  if we can find amongst the roots of  $E_7^{+++}$ one such   that $\Lambda =\sum_j p_j\lambda_j$.
Combining this with (3.2) and taking  the scalar product with $\lambda_i$  implies that
$$ n_k = l (\lambda_6,\lambda_k) -
\sum_jp_j(\lambda_j,\lambda_k) \ . \eqno(3.3)$$

Squaring the root $\alpha$ of (3.2) and applying the  knowledge of  the known lengths of roots in simply
laced Kac-Moody algebras one finds
$$
\alpha^2 = -{2\over 5}l^2 + \sum_{i,j} p_i(\lambda_i,\lambda_j) p_j   = 2, 0, -2, ... \eqno(3.4)$$

The Chevalley generator corresponding to the deleted  node 10 has 4  indices and  so a generator with  level
$l$ will have $4l$ $A_{D-1}$ indices. Any block of 10 fully anti- symmetrised indices will not transform
under the $A_9$ sub-algebra.  Using $m$ to denote the number of such blocks we  may write
$$\sum_{j=2}^9 p_j (10 - j) + 10 m = 4l \ .
\eqno(3.5)$$ Substituting this into (3.4) gives
$$
\alpha^2 = {1 \over 8} \left[ \sum_{j=1}^9 p_j^2 (10-j)(j-2) + 2 \sum_ {i,i<j} p_i p_j (10-j)(i-2) \right]
$$
$$
-{1\over 2} m  \sum_{j=1}^9  (10-j)p_j -{5\over 2} m^2= 2,0,-2,\dots  \eqno(3.6)
$$
We now show that any generator that satisfies  (3.5) will  automatically satisfy the root condition (3.3).
In particular, substituting (3.5)  into (3.3) we find that the  coefficients of the roots are given by
$$
n_j=\sum_{i<j} p_i(j-i)+ jm,\ j=1,\ldots, 6,\ \ n_7={3\over 4}(\sum_ {i=1}^9 (6-i)p_i+6m)+2p_9+p_8 \ ,
$$
$$
n_8={2\over 4}(\sum_{i=1}^9  (6-i)p_i+6m)+p_9,\ \ n_9={1\over 4}(\sum_ {i=1}^9 (6-i)p_i+6m) \ .  \eqno(3.7)
$$
We note that $4l-4\sum_ip_i -4m= \sum_{i=1}^9  (6-i)p_i+6m$. Also, the $n_j$ are all integers as required.

Dual generators are defined  to be those with no blocks of nine or  ten fully antisymmetrised indices.  This
condition may be written as
$$p_1 = m = 0 \ . \eqno(3.8)$$
Substituting this into (3.6) gives
$$
\alpha^2 = {1 \over 8} \left[ \sum_j^9 p_j^2 (10-j)(j-2) + 2 \sum_ {i<j}^9 p_i p_j (10-j)(i-2) \right] =
2,0,-2,\dots\eqno(3.9)
$$
This equation is independent of  $p_2$ and its terms are  positive .  Setting $p_2=0$ we find that the only
solutions are given by
$$ \eqalign {
\alpha_A = (0,0,0,0,0,0,0,0,0,1),\ \  p_6 = 1 \cr \alpha_B =  (0,0,0,1,2,3,2,1,0,2),\ \ p_3 = p_9 = 1 \cr }
\eqno(3.10)$$ where all other $p_i=0$. Solutions A and B correspond  to  the generators $R^{a_1 \dots a_4}$
and $R^{a_1\dots a_7, b}$ respectively. These have   multiplicity one.    The resulting low level field
content in the non-linear realisation is   a self dual four form $A_{a_1\dots  a_4}$ and the dual graviton
$A_{a_1\dots a_7, b}$ [9].

Setting  $p_2=1$ we find the solution
$$
\gamma = (0,0,1,2,3,4,3,2,1,2) \ .  \eqno(3.11)
$$
The generator corresponding to  this root has 0 multiplicity and so  it does not appear in the algebra.

The  roots corresponding to all possible dual generators in $E_7^{+++} $ can then be written in the form
$$ \eqalign {
\alpha_A(p_2) = \alpha_A + p_2\gamma\cr \alpha_B(p_2) = \alpha_B + p_2 \gamma\cr \alpha_C(p_2) = p_2
\gamma\cr} \eqno(3.12)$$ As in all the other cases, this corresponds to the  presence of all possible duality
transformations.


\vskip 1cm \noindent {\bf 4 \ \  ${D_{D-2}^{+++}}$}
\medskip
\noindent The nonlinear realisation of the  very extended $D_{D-2}^{++ +}$ algebras was conjectured as a
symmetry of a suitably extended low energy  effective action of the bosonic  string in D dimensions[20].  The
Dynkin diagram of ${D_{D-2}^{+++}}$ is given by

$$
\matrix{ & & & & & &0&D+1& &&& & & &0&D&\cr
      & & & & & &|& & & & & & & &|& &\cr
     0&-&0&-&0&-&0&-&0&.&.&.&0&-&0&-&0&\cr
     1& &2& &3& &4& &5& & & &D-3& &D-2& &D-1&\cr}
$$

For simplicity we consider $D\ge7$; the Dynkin diagram has a slightly  different structure for smaller $D$.
For the $D_{D-2}^{+++}$ series of algebras two nodes must be deleted to   give an $A_{D-1}$ algebra. The
simple roots of $D_{D-2}^{+++}$ are the simple roots of the $D_D$  algebra and   the simple root of the
deleted node $D+1$ which  may be written as
$$
\alpha_{D+1} = y - l_4 \eqno(4.1)$$ where  $l_4$ denotes a  fundamental weight of the $D_D$ algebra. Next we
delete node $D$ to find an $A_{D-1}$ gravity sub-algebra.  The simple  roots of $D_{D-2}^{+++}$ are now given
by the simple roots of the  $A_{D-1}$  sub-algebra,   the simple root of  node $D$, which we may write as
$$ \alpha_D = x - \lambda_{D-2}
\eqno(4.2)$$ and the simple root of equation (4.1). However,   we may  express the fundamental weight  $l_4$
of $D_D$ in terms of the fundamental weights of $A_{D-1}$  by [22]
$$
l_4 = \lambda_4 + {x \over x^2}(\lambda_{D-2}, \lambda_4) \eqno (4.3)$$ Noting that $\alpha_{D+1}^2 =
\alpha_D^2 = 2$  we find that $x^2 = {4\over  D}, y^2 = -2$.

A general root of $D_{D-2}^{+++}$ can be written
$$
\alpha = \sum_i n_i\alpha_i + l_x \alpha_D + l_y \alpha_{D+1} = l_y y  +  x (l_x-2l_y)- \Lambda, {\rm \
where\ }
$$
$$
\Lambda = l_y \lambda_4  + l_x\lambda_{D-2} -  \sum_i n_i\alpha_i  \eqno(4.4) $$ where there are now two
levels, $l_x$ and $l_y$  which refer to the deleted nodes $D$ and $D+1$  respectively. A highest weight
representation of $A_{D-1}$ with Dynkin indices $p_i $ can occur if the roots of the  Kac-Moody algebra
${D_{D-2}^{+++}}$ include the case when $\Lambda=\sum_i p_i\lambda_i$. Taking the  scalar product of both
sides with $\lambda_k$ gives
$$
n_k = l_y(\lambda_4,\lambda_k) + l_x (\lambda_{D-2}, \lambda_k) -
 \sum_j p_j(\lambda_j,\lambda_k) \eqno(4.5)$$

The square of the corresponding root is given by
$$
\alpha^2 = -2l_y^2 + {4\over D}\left( l_x - 2l_y \right)^2 + \sum_ {i,j} p_i(\lambda_i, \lambda_j) p_j=
2,0,-2,\dots \eqno(4.6) $$ where $\alpha^2$ has been constrained to  take the values given above as it
belongs to a Kac-Moody algebra [27].

The Chevalley generator corresponding to node $D$ has  2 $A_{D-1}$  indices, and the Chevalley generator
corresponding to node $D+1$ has $(D-4)$ indices. By  definition the multiple commutator of a generator with
level $ (l_x,l_y)$ contains the Chevalley generator corresponding  to node $D$ $l_x$ times,  adding $2l_x$
indices, and the Chevalley  generator corresponding to node $D+1$   $l_y$ times, adding $(D-4)l_y$ indices.
In total this gives $2l_x +  (D-4)l_y$  indices. Using $m$ to denote the  number of rank $D $ index blocks,
we may write
$$
\sum_j p_j(D-j) + Dm = (D-4)l_y + 2l_x \ . \eqno(4.7)$$ Substituting  this into (4.6) gives
$$ \alpha^2 = {1 \over D-2} \left[ 4d^2 + \sum_i p_i^2(D-2)(D-i)
+ 2 \sum_{i<j}p_i p_j(D-j)(i-2) \right]
$$
$$
-{4m \over (D-2)} \sum_i p_i(D-i)-{2m^2 \over (D-2)} = 2,0,-2,\dots  \eqno(4.8)
$$
In deriving this equation we have used the identity
$$
-Dl_y^2  + 2(l_x- 2l_y )^2 =
 {1 \over D-2} \left[ 2D(l_x-l_y)^2 -  \left( (D-4)l_y +
2l_x\right)^2\right] \eqno(4.9)$$ where $d=l_x-l_y$.

In fact,  any solution that satisfies equation (4.7) will  automatically satisfy the root condition of
equation (4.50 and  one  finds that
$$
n_j=\cases{\sum_{i,i<j} p_i(j-i) +mj, \ \ j=1,2,3,4\cr \sum_{i,i<j} p_i (j-i) +mj+l_y(j-4) ,\ \ j=4,\dots
,D-1\cr }
$$
$$
n_{D-1}= l_y+l_x-\sum_ip_i-m \eqno(4.10)$$

Dual generators are defined to be those with no blocks of $D$  or $D-1 $ fully anti-symmetrised indices. This
may be written as $ p_1 = m = 0$ which when  substituting this into  equation (4.8) gives
$$
\sum_j p_j(D-j) = (D-4)l_y + 2l_x \ . \eqno(4.11)
$$
Equation (4.8) then  becomes
$$
\alpha^2 = {1 \over D-2} \left[ 4d^2 + \sum_j p_j^2(D-2)(j-2) + 2  \sum_{i<j}p_ip_j (D-j)(i-2) \right] =
2,0,-2,\dots \eqno(4.12) $$ This equation is independent of $p_2$ and  all terms in the middle equation are
positive definite.   The general solution to this equation can be  found following similar arguments to that
deployed for the case of $A_{D-3}^{+++}$, in particular below  equation (2.14).

Setting $p_2=0$ we find the following roots
$$ \eqalign{
\alpha_A &= (0,\dots,0,1,0),\ \  p_{D-2} = 1,  \cr \alpha_B &= (0, \dots,0,1),\ \  p_4=1, \cr \alpha_C &=
(0,0,0,1,\dots,1,0,1,1),\ \  p_3 = p_{D-1} .\cr } \eqno(4.13)$$ All  other $p_i$ are 0. The generators
corresponding  to the roots A,  B and C  are  $R^{a_1 a_2}$, $R^{a_1 \dots a_{D-4}}$  and $R^{a_1 \dots
a_{D-3}, b}$ respectively. These all have multiplicity one.

Setting $p_2=1$ we also find the solution
$$
\gamma = (0,0,1,2,\dots,2,1,1,1) \eqno(4.14)
$$
corresponding to the generator $R^{a_1 \dots a_ {D-2}}$ which has  multiplicity one.

As a result,  the non-linear realisation contains at low levels the  fields of gravity and a dilaton, as the
rank of ${D_{D-2}^{+++}}$ is one greater than that $D$ the dimension  of space-time, a two form $A_{a_1
a_2}$, its dual  $A_{a_1\dots a_{D-4}}$,  the dual graviton $A_{a_1\dots a_ {D-3}, b}$ and the field $A_{a_1
\dots a_{D-2}}$, which is the dual of the dilaton. These are the on-shell  states of the effective action of
the bosonic string generalised to $D$ dimensions [20,9].

The roots corresponding to all dual generators may be  written in the  form
$$ \eqalign{
\alpha_A(p_2) &=  \alpha_A+p_2\gamma  \cr \alpha_B(p_2) &= \alpha_B +p_2\gamma  \cr \alpha_C(p_2) &=
\alpha_C+p_2\gamma  \cr \alpha_D (p_2)&=p_2\gamma  \cr} \eqno(4.15)$$

\vskip 1cm \noindent {\bf {5 \ \  $E_6^{+++}$}}
\medskip
\noindent At low levels the nonlinear realisation of the very extended $E_6$ algebra has precisely the field
content [9] of the oxidation  endpoint of the non-supersymmetric $E_6$ coset theory described in [28].  The
Dynkin Diagram of ${E_6^{+++}}$ is given by

$$
\matrix{
    & & & & & & & & &0&9&     \cr
    & & & & & & & & &|&&     \cr
    & & & & & & & & &0&8&     \cr
        & & & & & & & & &|&   \cr
        &0&-&0&-&0&-&0&-&0&-&0&-&0& \cr
        &1& &2& &3& &4& &5& &6& &7& \cr}
$$
As usual nodes one, two and three are the very extended, over  extended and affine nodes. In the case of
$E_6^{+++}$ node 8 is deleted leaving  the algebra $A_7 \otimes A_1$.  This decomposition is similar to the
decomposition of $E_{11}$   which is appropriate to IIB supergravity  [8,9].  The roots of $E_6^{+++}$ can be
written as the  roots of  $A_7$, $\alpha_i, i = 1\dots7$, the root of  the $A_1$ algebra $ \beta$, and the
deleted root $\alpha_8$, which may be written
$$ \alpha_8 = x - \lambda_5 - \mu
\eqno(5.1) $$ where $\lambda_5$ is a fundamental weight of the $A_7$  algebra and $\mu$ is the fundamental
weight of the $A_1$ subalgebra. Knowing $\alpha_8^2 = 2$, we find that $x^2  = - {3\over8}$.  A general root
of ${E_6^{+++}} $ may be written as
$$
\alpha = \sum_{i=1}^7 n_i \alpha_i + l \alpha_8 + r\beta = lx -  \Lambda \ \ {\rm where } \ \ \Lambda =
l\lambda_5 + l\mu - \sum_{i=1}^7 n_i\alpha_i - r\beta \ . \eqno(5.2) $$ A highest weight representation of
$A_7 \otimes A_1$ with Dynkin indices $p_i,q$ can occur if the ${E_6^{+++}} $ include the possibility
$\Lambda = \sum_ip_i\lambda_i + q\mu$. Dotting this with $\lambda_k$ and $\mu$  in turn gives the pair of
equations
$$\eqalign{
n_k &= l (\lambda_5,\lambda_k) - \sum_{j=1}^7 p_j (\lambda_j, \lambda_k) \cr r &= {l-q \over 2} \cr }
\eqno(5.3)$$ Note that $(\mu, \mu) = 1/2$, and $(\beta, \mu) = 1$.    Squaring the expression for $\alpha$ in
(5.2) gives
$$
\alpha^2 = -3/8 \ l^2 + \sum_{i,j} (\lambda_i,\lambda_j) p_i p_j+ q^2/2 \ .  \eqno  (5.4)
$$

The Chevalley generator corresponding to node  8 has three $A_7$   indices. As a result, a level $l$
generator, the multiple commutator for which by definition contains the  Chevalley generator corresponding to
node  8 $l$ times, will have $3l$ $A_7$ indices in total.  We note that the   Chevalley generator
corresponding to node 9 has no $A_7$ indices, so it does not contribute to the index count. As a  result
$$ \sum_jp_j(8-j) + 8m=3l
\eqno(5.5)$$ where $m$ denotes the number of blocks of 8 fully  antisymmetrised  indices.

Substituting for $l$ from (5.5) into (5.4) gives
$$
\alpha^2 = {1 \over 6} \left( \sum_j p_j^2(8-j)(j-2) + 2 \sum_{i<j}  p_ip_j(8-j)(i-2) - 16m^2 -
4m\sum_jp_j(8-j) \right) +{1\over 2}q^2
$$
$$
= 2,0,-2,\dots\eqno(5.6)
$$

We now show that any generator that satisfies (5.5)  automatically
 satisfies the root condition (5.4).
Substituting (5.5) into (5.4) we  find
$$
n_j=\cases{\sum_{i ,i<j} p_i(j-i)+jm,\ \ \  j\le 5\cr n_6= {2\over 3} (\sum_i(5-i) p_i+5m) +p_7  \cr n_7=
{1\over 3}(\sum_i(5-i) p_i+5m)   \cr}  \eqno (5.7)
$$
We note that $3l-3\sum_i p_i-3m =(\sum_i(5-i) p_i+5m$ and   the $n_j $ are positive integers as required.

Dual generators are defined to be those with  no blocks of 7 or 8   fully antisymmetrised indices.  This may
be written as
$$ p_1 = m = 0 \ . \eqno(5.8) $$
Substituting these into (5.6) gives
$$ \alpha^2 = {q^2 \over 2} + {1 \over 6} \sum_j p_j^2(8-j)(j-2)
+ {1 \over 3} \sum_{i<j} p_i p_j (8-j) (i-2) = 2,0,-2,\dots \eqno(5.9) $$ The middle terms in this equation
are positive definite and $p_2$  is undetermined. Setting $p_2 =0$ gives the solutions
$$ \eqalign{
\alpha_A &= (0,0,0,0,0,0,0,1,0),\ \  q=1, p_5 = 1 \cr \alpha_B &=  (0,0,0,1,2,1,0,2,1), \ \ q=0, p_3 = p_7 =
1 \cr \alpha_C &= (0,0,0,0,0,0,0,0,-1), \ \ q=2 \cr } \eqno(5.10) $$  All other  $p_i$ are zero. Generators
with $q=0$, $q=1$ and $q=2$ are  singlets, doublets, and triplets of $SL (2,R)$ respectively. We note that
the coefficients of the expression  for a general  root of equation (5.3)  in terms of simple roots
corresponding to roots 8 and 9 are $l$ and $r$ respectively. The generators  corresponding to the roots A and
B  are $R^{a_1a_2a_3, \alpha}$ and $R^{a_1\dots a_5, b}$ respectively, where  $\alpha,\beta=1,2$ indices
label the vector representation of $SL(2,R)$. Root C corresponds to the  generators $R^{(\alpha\beta)}$ which
are none other than the generators of $A_{1}$ arising from node nine. Unlike  the other representations of
$A_7\otimes A_1$ mentioned above  which occur entirely within the negative level  root space of ${E_6^{+++}}$
with a copy in the positive root space this representation contains  positive and  negative levels. This is
related to the fact that it belongs to the adjoint representation of $A_7\otimes A_1 $ with respect to which
we are decomposing the ${E_6^{+++}}$ algebra. Following the derivation of equation (5.2)  and reading
reference [26] one sees that in this case a negative value of $r$ is allowed.

If $p_2=1$ we find the two solutions
$$
\eqalign{ \alpha_D &= (0,0,1,2,3,2,1,2,1), \ \ q=2 \cr \gamma &=  (0,0,1,2,3,2,1,2,1), \ \ q=0 \cr }
\eqno(5.11)
$$
The corresponding generators are $R^{a_1\ldots a_4,(\alpha\beta)}$  and $R^{a_1\ldots a_4}$ and these have
multiplicity one and zero respectively.

The fields corresponding to the above generators are two scalars $\phi $ and $\chi$ that belong to the coset
$SL(2,R)$ with  local sub-algebra $SO(2)$ associated with root C,  their duals $A_{a_1\ldots
a_4,(\alpha\beta)}$ which are subject to one constraint and are associated with root D, a  doublet of three
forms $A_{a_1a_2a_3,\alpha}$, whose four form field strength is self-dual  and are associated with root A and
finally the dual graviton $A_{a_1\dots a_5, b}$ associated with root  B. This is the content of an eight
dimensional non-supersymmetric theory.

The roots of all dual generators may be written in the form
$$ \eqalign{
\alpha_A(p_2) &= \alpha_A+ p_2 \gamma \cr \alpha_B(p_2) &= \alpha_B +p_2 \gamma  \cr \alpha_C(p_2) &=
\alpha_C+p_2 \gamma  \cr \alpha_D(p_2) &= \alpha_D+p_2 \gamma \cr  \alpha_E(p_2) &= p_2 \gamma \cr}
\eqno(5.12)
$$

In Appendix A we derive all the forms arising from dimensional reduction to 4 dimensions and above.

\vskip 1cm \noindent {\bf {6 \ \  $G_2^{+++}$}}
\medskip
\noindent The nonlinear realisation of $G_2^{+++}$ contains at low levels the fields of $N=1$ five
dimensional supergravity [9]. The Dynkin Diagram of  $G_2^{+++}$ is given by

$$ \matrix {
&&&&&&0&&5 \cr &&&&&&{|||}&& \cr &&&&&&\widehat{|||}\cr 0&-&0&-&0&-&0  \cr 1&&2&&3&&4 \cr } $$

The simple root vectors can be  chosen to be
$$\matrix {
(\alpha_5, \alpha_5) = 2/3 ,\ \ (\alpha_i,\alpha_i)=2, \  i = 1,\dots , 4,\cr (\alpha_i,\alpha_j)=-1,\ i,j =
1,\dots ,4,\  |i-j| = 1 \cr (\alpha_5,\alpha_4) = -1\cr } \eqno(6.1) $$ All other scalar products are 0.

In this case, the node labelled 5 is deleted to leave an $A_4$  gravity subalgebra. The deleted node can be
written as
$$
\alpha_5 = x - \lambda_4 \ . \eqno(6.2)
$$
Given $\alpha_5^2=2/3$ we find $x^2 = - 2/15$.  A general root of $G_2^{+++}$ may be written
$$
\alpha = \sum_{i=1}^4 n_i\alpha_i + l\alpha_5  = lx - \Lambda, \ \ \  {\rm where }\ \ \ \Lambda = l\lambda_4
- \sum_i n_i\alpha_i \ . \eqno(6.3)
$$
A highest weight representation of $A_4$ with Dynkin  indices $p_i$  can occur if we can choose $\Lambda
=\sum_i p_i\lambda_i$.  Dotting this  with $\lambda_k$ we find
$$
n_k = l(\lambda_4,\lambda_k) - \sum_j p_j(\lambda_j,\lambda_k) \eqno  (6.4)$$

Squaring the expression for $\alpha$ in (6.3) gives
$$
\alpha^2 = -{2\over 15}l^2  + \sum_{i,j}p_ip_j(\lambda_i,  \lambda_j)  = {1\over
3}(6,2,0,-4,-10,-16,-22,\dots) \eqno(6.5)
$$
Where $\alpha^2$ is constrained to take the values above as it  belongs to the  Kac-Moody algebra
$G_2^{+++}$ [27].   The Chevalley generator corresponding to node 5 has one $A_4$  index, so a level $l$
generator, the multiple commutator for which contains the generator  corresponding  to node 5 $l$ times will
have $l$ $A_4$ indices in total.  As a result
$$
l = \sum_j(5-j)p_j + 5m \eqno(6.6)
$$
where $m$ denotes the number of blocks of 5 fully  anti-symmetrised  indices. Substituting this into (6.5)
gives
$$
\alpha^2 = 1/3\left(\sum_j^4 p_j^2 (5-j)(j-2) + 2\sum_{i<j}^4p_ip_j(5- j)(i-2) - 4m\sum_j^4(5-j)p_j - 10m^2
\right) \eqno(6.7)
$$
We now show that any generator that satisfies  (6.6) possesses a root  that automatically satisfies the root
condition (6.4).   Substituting (6.6) into (6.4) we find that
$$
n_j = \sum_{k<j}p_k(j-k) + mj \ . \eqno(6.8)
$$
Note that it  gives positive integer values for the root components  $n_k$.

Dual generators are defined to be those with no  blocks of 4 or 5  totally anti\-symmetrised indices, which
may be written as
$$
m = p_1 = 0 \ . \eqno(6.9)
$$
Substituting these into (6.7) gives
$$
\alpha^2 = {1\over 3}\left(\sum_j p_j^2 (5-j)(j-2) + 2\sum_{i<j}p_ip_j(5-j) (i-2) \right)
$$
$$
= {1\over 3}(6,2,0,-4,-10,-16,-22,\dots) \eqno(6.10)
$$
Note that $p_2$ is absent from this formula, and the  terms  in the  middle equation are positive.

Taking $p_2=0$ in equation (6.10) we get the following solutions
$$ \eqalign{
\alpha_A &= (0,0,0,0,1), p_4 = 1 \cr \alpha_B &= (0,0,0,1,2), p_3 = 1  \cr \alpha_C &= (0,0,0,1,3), p_3 = p_4
= 1\cr } \eqno(6.11) $$ All other $p_i$ are 0. The roots $\alpha_A$ and  $\alpha_B$ correspond to the
generators $R^a$ and $R^{a_1a_2}$ respectively.  These give rise to the   1-form  $A_{a}$ and 2-form
$A_{a_1a_2}$  in the nonlinear realisation. Since the theory is in five dimensions the  latter is the dual
field of the former.  The root $\alpha_C$ corresponds to  the generator $R^{a_1 a_2, b}$.  The
corresponding field is the  dual graviton. All these fields have multiplicity one. Thus at lowest  levels the
non-linear realisation contains gravity and a vector field as its on-shell degree of freedom [9].  This is
five dimensional $N=1$ (eight supercharges) supergravity. This theory was constructed in [29,30] and is
sometimes referred to as $N=2$ as it gives an $N=2$ theory when dimensionally reduced.

Setting $p_2=1$ gives the solution
$$
\gamma = (0,0,1,2,3) \eqno(6.12)
$$
This root vector  has multiplicity 0 so does not contribute a field  to the non linear realisation.

One finds that all dual generators have the roots
$$ \eqalign {
\alpha_A(p_2) &= p_2 \gamma + \alpha_A \cr \alpha_B(p_2) &= p_2  \gamma + \alpha_B \cr \alpha_C(p_2) &= p_2
\gamma + \alpha_C \cr \alpha_D (p_2) &= p_2 \gamma\cr } \eqno(6.13) $$

We will now find the space-filling and next to space-filling forms of the $G_2^{+++}$ theory in five, four
and three dimensions in the same way as was done for the $E_{8}^{+++}$ theory in reference [16]. In five
dimensions, that is the highest dimension the theory can exist, and the one considered above, we find that  a
three form generator appeared  in equation (6.12),  but this had multiplicity zero. Examining equation (6.7)
we also find a solution with $p_1=1, m=0$ which corresponds to  a four form  generator, but this also has
multiplicity zero [9,31]. However, there  are no scalars in this theory and so there are  no fields to which
the field strength of the three form can be dual. As such it is not required. That there is no  four form
implies that there are no deformations and so no gauged supergravities that arise  from the field strength of
the four form. There is a known gauging of this theory [32]. However it gauges a $U(1)$ sub-algebra of a
$USp(2)$ symmetry that does not act on the bosonic fields,  but only on the gravitino. Hence one does not
expect the gauged theory to be associated with a deformation of the bosonic sector. To predict this gauged
theory from the non-linear realisation one must first introduce the gravitino as it is only on this field
that the $USp(2)$ symmetry acts.

Let us consider adding  a three form and four form  gauge fields and extending the known supersymmetry. Given
that there is no internal symmetry in the five-dimensional $G_2^{+++}$ non-linear realisation, the only
allowed supersymmetry variation of the three form is given by $\delta A_{a_1a_2a_3}= i\bar
\epsilon^i\gamma_{[a_1a_2} \psi_{a_3]}^j\epsilon_{ij}$. However, taking the commutator of two supersymmetry
transformations we find that the supersymmetry algebra does not close and so one should not introduce a three
form. The same conclusion holds for the four form. This is consistent with the fact that $G_2^{+++}$ assigns
multiplicity zero to these fields.

We now consider the $G_2^{+++}$ theory in four dimensions. This can be found by just dimensionally reducing
the theory in five dimensions or deleting node four of the  $G_2^{+++}$ Dynkin diagram and analysing the
content with respect to the remaining $A_3\otimes A_1$ algebra. The latter $A_1$ factor is the internal
symmetry group of the four dimensional theory. The results are the same, but we will consider the former
approach.  Using the tables of  generators of references [9,31], we find the fields; two scalars ($h_5{}^5,
A_5$) which belong to the coset $SL(2,R)$ with local sub-algebra $SO(2)$, an $SL(2,R)$ quadruplet of vectors
($h_a{}^5, A_{a }$, $ A_{a 5},A_{a 5,5} $) which satisfy self-duality conditions,  a triplet of two forms
($A_{a_1a_2}, A_{a_1a_2,5}, A_{a_1a_2,5,5}$) which are subject to one constraint and are dual to the two
scalars. We also find a doublet of three forms ($A_{a_1a_2a_3,5}, A_{a_1a_2a_3 5,5}$) and finally a triplet
four forms  ($A_ {a_1a_2a_3a_4,5}, A_{a_1 a_2 a_3 a_4 5,5}, A_{a_1 a_2 a_3 a_4 5,5,5}$).

This theory is $N=2$ supergravity coupled to one $N=2$ vector multiplet which was constructed in [30] and
more recently in [33]. However, the $G_2^{+++}$ formulation  is a democratic formulation in that it includes
the Hodge duals of all the field strengths. As it possesses a doublet of four forms  it predicts  two
possible deformations that is gauged supergravities. The non-linear realisation and the supersymmetry closure
of this theory is under investigation and the preliminary results confirm the $G_2^{+++}$ picture [34].

We now turn to the three dimensional $G_2^{+++}$ theory  which results from deleting node three of the
$G_2^{+++}$ Dynkin diagram leaving a $G_2$ internal symmetry.  We will compute the forms by dimensionally
reducing the five dimensional theory. We find six scalars ($h_i{}^j, A_{i}, A_{ij}$) which belong to the
non-linear realisation $G_2$ with local sub-algebra $SU(3)$. There are fourteen vectors ($h_a{}^i, A_{a }$,
$A_{ai}, A_{ai,j}$, $ A_{aij,k}, A_{aij,kl}$, $A_{aij,kl, h}$).  These all have multiplicity one and belong
to the adjoint representation of $G_2$ and are dual to the scalars  taking into account that they   satisfy
some constraints.  The two forms belong to the 27 dimensional representation of $G_2$ and an additional
singlet of $G_2$ (in the original version of this paper, the singlet deformation was erroneously missing, see
the end of appendix B for details). In fact to find these latter forms one has to go slightly beyond the
tables of references [9,31] and take into account the fact that some generators have multiplicity two. Hence,
this theory possess a set of deformations parameterised by the 27 dimensional representation of $G_2$ and a
singlet of $G_2$, and  a corresponding set of gauged supergravities.

In reference [35] it is pointed out that the  $G_2^{+++}$ theory does not satisfactorily encompass the known
gauged supergravity of Ref. [32]. We believe that the reason for this mismatch is due to the fact that in
[32] it is a $U(1)$ subgroup of the fermionic symmetry $USp(2)$ that is gauged, and given that there is no
$USp(2)$ inside $G_2^{+++}$ in five dimensions, this result is totally expected. Introducing the fermions in
the non-linear realisation might lead to a solution of this problem. We will discuss this in more detail in
appendix C.

\vskip 1cm \noindent {\bf {7 \ \  $F_4^{+++}$}}
\medskip
\noindent The Dynkin diagram of $F_4^{+++}$ is given by
$$ \matrix {
&&&&&&&&&0&7 \cr &&&&&&&&&|\cr &&&&&&&&&0&6\cr &&&&&&&&&\Uparrow \cr  &0&-&0&-&0&-& 0& -& 0\cr
&1&&2&&3&&4&&5\cr } $$

Given the Cartan matrix we may choose the  root vectors to obey
$$ \matrix {
(\alpha_i,\alpha_i)=2, i= 1, \dots, 5\cr (\alpha_6,\alpha_6)= (\alpha_7,\alpha_7)=1 \cr
(\alpha_i,\alpha_j)=-1,\ \  i,j = 1, \dots, 6, |i-j|=1 \cr (\alpha_6,\alpha_7)=-1/2 \cr } \eqno (7.1) $$ All
other scalar products are 0. In the case of $F_4^{+++}$  the node labelled 6 is deleted, leaving  an $A_5$
gravity algebra, and an $A_1$ algebra.  The roots of $F_4^{+++}$ can be written as the roots of $A_5 $,
$\alpha_i, i=1,\dots, 5$, the root of $A_1$, $\beta$, and the deleted  root $\alpha_6$. The deleted root  may
be written as
$$
\alpha_6 = x - \lambda_5 - \mu \eqno(7.2)
$$
where $\lambda_5$ is a fundamental weight of the $A_5$ subalgebra, and $\mu$ is the fundamental weight of the
$A_1$ subalgebra. The root of the $A_1$ subalgebra, $\beta$, is  normalised to have length 1, hence
$\mu^2=1/4, \mu\beta = 1/2$.  We find $x^2 = -1/12$.

A general root of $F_4^{+++}$ can be written
$$
\alpha = l\alpha_6 + \sum_{i=1}^5 n_i\alpha_i + r\beta = lx-\Lambda \  \ {\rm where} \ \ \Lambda = l\lambda_5
+ l\mu - \sum_i n_i\alpha_i - r\beta \eqno(7.3)$$ A highest weight  representation of $A_5 \otimes A_1$ with
Dynkin indices $p_i, q$ respectively can occur if we can choose $\Lambda =  \sum_i p_i \lambda_i + q\mu$.
Dotting this with $\lambda_k$ and $\mu$ in turn gives the pair of equations
$$ \eqalign {
n_k &= l(\lambda_5,\lambda_k) - \sum_i (\lambda_i, \lambda_k) \cr r  &= {l - q \over 2} \cr } \eqno(7.4) $$
Squaring the expression for $\alpha$ in (7.3) gives
$$
\alpha^2 = -{1\over 12}l^2 + \sum_{i,j}p_ip_j(\lambda_i,\lambda_j) +  q^2/4 = 2,1,0,-1,\dots \eqno(7.5)$$
where $\alpha^2$ is constrained to take the values above.

The Chevalley generator corresponding to node 6 has one $A_5$ index,   so a level $l$ generator, the multiple
commutator for which contains the generator corresponding to node 6 $l $ times, will have $l$ $A_5$ indices
in total.  The Chevalley generator corresponding to node 7 has no $A_5$  indices, so it does not contribute
to the index total. As a result
$$
l = \sum_j (6-j) p_j + 6m \eqno(7.6)
$$
where $m$ denotes the number of blocks of 6 fully antisymmetrised   indices. Substituting this into (7.5)
gives
$$
\alpha^2 = {1\over 4}\sum_{j=1}^5p_j^2(6-j)(j-2) + {1\over 2} \sum_ {i<j}^5 p_i p_j(6-j)(i-2) + {q^2\over 4
}- m\sum_j(6-j)p_j - 3m^2
$$
$$
= 2,1,0,-1,\dots \eqno(7.7)
$$
We now show that any generator that satisfies (7.6)  automatically  satisfies the root condition (7.4).
Substituting (7.6) into (7.4) we  find
$$
n_j = \sum_{i<j}p_i(i-j) + mk,\ \  j=1, \ldots ,5 \ . \eqno(7.8) $$ We  note that $n_7 = r$, and $n_6 = l$.

Dual generators are defined to be those with no blocks of  5 or 6 fully anti-symmetrised $A_5$ indices. This
may be written as
$$
m = p_1 = 0 \ . \eqno(7.9)
$$
Substituting these into (7.7) gives
$$
\alpha^2 = {1\over 4}\sum_{j=1}^5p_j^2(6-j)(j-2) + {1\over 2} \sum_ {i<j}^5p_ip_j(6-j)(i-2) + {q^2\over 4 }=
2,1,0,-1, \dots \eqno(7.10)$$

We now find all the solutions to equation (7.10). The middle terms of this equation are positive definite
with $p_2$ undetermined. Taking $p_2=0 $ we get the following solutions for $\alpha^2 = 1$
$$ \eqalign {
\alpha_A &= (0,0,0,0,0,1,0), p_5=1, q=1 \cr \alpha_B & =  (0,0,0,1,2,3,1), p_3=1, q=1 \cr \alpha_C &=
(0,0,0,0,1,2,1), p_4=1, q=0 \cr \alpha_D &= (0,0,0,0,0,0,-1), q=2  \cr } \eqno(7.11)$$ All other $p_i$'s are
zero. The $p_i$s and $q$ are the highest weight components of  the $A_5\otimes A_1$ representation.
Generators with $q=0, 1, 2$ are $A_1$ singlets,  doublets and triplets  respectively.

The generators corresponding to roots A B, C and D are $R^{a ,\alpha} $, $R^{a_1 a_2 a_3,\alpha}$ and $R^{a_1
a_2}$ and $R^{(\alpha\beta)}$ respectively, where   $\alpha, \beta=1,2 $. These have multiplicity one except
for $R^{a_1 a_2}$  which has multiplicity zero. The last generator is  symmetric in $\alpha\beta$ and is
just the triplet of  generators  of $A_1$ itself. These are at level zero with  respect to $n_6$ and contains
generators at levels $n_7=0\pm 1$. The appearance of a negative level is allowed  for the same reason as it
did in the case of $E_6^{+++}$.  These generators give rise to the fields expected  in  the non-linear
realisation except for the multiplicity zero  generator which  leads to no field and the  generator
$R^{(\alpha\beta)}$ which only gives rise to two scalars $\phi$ and $\chi$ due to the   local symmetry  which
being the Cartan involution invariant sub-algebra  includes the $SO(2)$ part of $A_1$.

The solutions for  $\alpha^2 = 2$, still with $p_2=0$, are given by
$$ \eqalign {
\alpha_E &= (0,0,0,0,1,2,0),\ \  p_4=1, q=2 \cr \alpha_F &=  (0,0,0,1,2,4,2),\ \  p_3=p_5=1, q=0\cr }
\eqno(7.12) $$ These both have multiplicity one and correspond to the generators $R^ {a_1a_2,(\alpha\beta)}$
and $R^{a_1a_2a_3,b}$. If we take $p_2=1$ we find the generators
$$ \eqalign {
\alpha_G &= (0,0,1,2,3,4,1),\ \  p_2=1, q=2  \cr \gamma & =  (0,0,1,2,3,4,2),\ \  p_2=1, q=0 \cr }
\eqno(7.13) $$ The second generator has multiplicity zero while the first has  multiplicity one and
corresponds to the generator $R^{a_1\ldots a_4, (\alpha \beta)}$.

Up to the levels considered the field content of the non-linear  realisation is: two scalars $\phi,\chi$ that
belong to the coset $SO(1,2)$, with local sub-algebra $SO(2)$,   together with  their duals $A_{a_1\ldots
a_4,(\alpha\beta)}$ (which are subject to one constraint), two  vectors $A_{a,\alpha}$ in the spinor
representation and their duals  $A_{a_1a_2a_3,\alpha}$, two  self- dual two forms and one anti- self dual two
form and the dual graviton. This makes up (1,0) supergravity in six  dimensions ( $h_a{}^b$ $A_{a_1a_2}^-$)
and coupled to two  vector multiplets ($A_{a,\alpha}$) and two tensor  multiplets ($A_{a_1a_2}^+$ and
$\phi,\chi$) and well as the dual graviton $R^{a_1a_2a_3,b}$ [9]. The $\pm$ on the  two forms indicate their
self-duality and we have not shown their $SO(1,2)$ indices, but they  combine together  to form the vector
representation of $SO(1,2)$.

The roots of all dual generators may be written in the form
$$ \eqalign {
\alpha_I(p_2) &=  \alpha_I+p_2 \gamma, \ \ I=A,B,C,D,E,F,G \cr  \alpha_H(p_2) &=  p_2 \gamma \cr} \eqno(7.14)
$$ Thus, as with all the other cases, we find the non-linear realisation  contains all possible dual
descriptions of the on-shell degrees of freedom of the theory.

In Appendix B we derive all the forms resulting from the $F_4^{+++}$  non-linear realisation in 4 dimensions
and above.
\vskip 1cm \noindent {\bf {8 \ \ Summary of Results and Discussion }}
\medskip
\noindent In this paper we have considered  all the very extended algebras $G^{+++}$ with the exception of
the $B$ and $C$ series. We have deleted one, and in some cases two nodes, from the Dynkin diagram to find a
preferred $A_{D-1}$ algebra, with in two cases an additional $A_{1}$ algebra,   and we have analysed the
content of the algebra $G^{+++}$ in terms of this $A_{D-1}$ algebra. We have shown  that all the generators
of $G^{+++}$ can be written with a set of  $A_{D-1}$ indices the total number of which  obeys a  condition
that depends on the level, or levels, of the generator in question. We found that this level  matching
condition automatically solves, in all cases,  the condition that a highest weight representation of $A_
{D-1}$ occurred amongst the root space of $G^{+++}$. As a result, necessary conditions for the roots of the
Kac- Moody algebra $G^{+++}$ become  the condition on the length of the root squared and the level matching
condition itself. Despite the fact that the level matching conditions varied from algebra to algebra the
condition on the length of the root squared   has a universal form given by
$$ \alpha^2 =  {1 \over (D-2)} \left( \sum_j p_j^2 (D-j)(j-2) + 2
\sum_{i<j} p_ip_j(D-j)(i-2) \right)
$$
$$
-{4m\over (D-2)} \sum_i(D-i)p_i -{2m^2 D\over (D-2)}+c{q^2\over 4} +{4d^2\over (D-2)} = 2,\dots \eqno(8.1)$$
The constant $c$ relates to the presence of an additional  $A_1$ algebra  that survives the deletion and it
is zero for all cases except for $E_6^{+++}$ and $F_4^{+++}$ where it is 2  and 1 respectively. The  integer
$d$ is zero for all cases except for $D_{D-2}^{+++}$ where it is the difference  in the two levels. For the
case of a symmetric Kac-Moody algebra $ \alpha^2 $ can only take the values  $2,0,-2,\dots $ for the
non-symmetric case the possible values are given in this paper.

Consequently,  the necessary condition for a representation of $A_ {D-1}$ with Dynkin indices $p_j$ and $m$
blocks of $D$ totally anti-symmetrised indices to occur in the  algebra $G^{+++}$  is the condition of
equation (8.1) and the level matching condition.  In all cases,  we have also found a formula for the
$G^{+++}$ roots in terms of their Dynkin indices $p_j$ and $m$. Indeed, we can think of  the set of integers
$\hat p_{\hat j}=(m,p_j)$ as belonging to a lattice and equation  (1.13) as  containing the  scalar product
on this lattice. The possible  generators of the Kac-Moody algebra   $G^{+++}$  correspond to   points $\hat
p_{\hat j}$ of the lattice that have an allowed length squared and obey the rather trivial level matching
condition. This, and the physical nature of the higher level fields, suggests that it  may not be totally
impossible to list  what are the generators of these very extended Kac-Moody algebras.

We then defined the notion of dual generators which are those that  have no blocks of $D$ and $D-1$ totally
anti-symmetrised indices, that is $m=0=p_1$. Substituting these  conditions into equation (8.1) we find it
simplifies, all  the terms are positive and we were able to find all possible solutions. As is apparent from
equation (8.1) we always have a class of solutions with $p_2$ taking  any positive integer value all other
$p_j=0$. It is  straightforward to verify that this always satisfies the level matching condition.  The
solutions with $p_2=0$ correspond in the non-linear realisation to a  set of  fields that give the simplest
description  of  the on-shell degrees of freedom of the theory   together with a set of dual fields whose
fields strengths are related to those of the original fields. Thus one finds  in this sense a democratic
formulation. The only other dual generators are all the just mentioned solutions  but with the addition of
the Dynkin index $p_2$ which can take any positive integer value. In the non-linear  realisation this
corresponds to adding blocks of $D-2$ totally antisymmetric   indices. These encode all possible  ways of
writing the on-shell degrees of freedom of the theory and their presence means that  the theory  encodes all
possible duality transformations of these on-shell degrees of freedom. With the assignment of indices to the
generators given in this paper, the generators of the affine subalgebra $G^+$ are just given by  taking the
indices to only take the values $0, \ldots D-3$. In this  case,  all generators with blocks of $D-1$ and $D$
indices are absent and one is left with the dual generators with the restricted  index range. Indeed, the
existence of the dual generators can be seen as a consequence of the affine subalgebra. One can think of the
role  of the dual generators as lifting the infinite number of duality  relations found in two dimensions up
to $D$ dimensions.

We have  explained that  the $G_2^{+++}$ theory in five dimensions does not possess  three and four form
fields and why  this is compatible with the known gauged supergravity theory, contrary to that claimed in
reference [35]. We have also computed the form fields for the $G_2^{+++}$ theory in three and four dimensions
and predicted the corresponding deformations, or gauged supergravities. A detailed analysis of all the forms
arising in lower dimensions for the cases of $E_6^{+++}$ and $F_4^{+++}$ is performed in two separate
appendices.

\vskip 1cm \noindent {\bf {Note added }}
\medskip
The referee has asked us to comment on the occurrence of an $SL(2)$  doublet and an $SL(2)$ quadruplet of
10-forms in the $E_{11}$ formulation of the IIB theory. Although these appeared for the first time in the
tables of Ref. [9], it was in [36] that it was stressed that $E_{11}$ predicts that the RR 10-form of IIB
belongs to a quadruplet. This prediction was considered unexpected and therefore was assumed to signal a
failure of the $E_{11}$ non-linear realisation [36]. It was only later that a doublet and a quadruplet of
10-forms were shown to occur in the IIB supersymmetry algebra [12], thus giving a highly non-trivial check of
the $E_{11}$ predictions.

\vskip 1cm \noindent {\bf Acknowledgments}
\medskip
\noindent We would like to thank Axel Kleinschmidt for discussions on affine algebras. Peter West would like
to thank  Andrew Pressley for  discussions on group theory. The research of Peter West was supported by a
PPARC senior fellowship PPA/Y/S/2002/001/44.  The work of Fabio Riccioni and Peter West is also supported by
a  PPARC rolling grant PP/C5071745/1 and  the  EU Marie Curie, research training network grant
HPRN-CT-2000-00122. The research of Duncan Steele is supported  by a PPARC Ph.D fellowship.

\vskip 1cm \noindent {\bf A \ \ $E_6^{+++}$ in lower dimensions}
\medskip
\noindent In this Appendix we determine all the forms that arise in the non-supersymmetric $E_6^{+++}$
non-linear realisation in dimensions from 8 to 4, with the exception of the 4-forms in 4-dimensions. The
scalar manifold is $Sl(2)/SO(2)$ in 8 dimensions, $Sl(2)/SO(2) \times R^+$ in 7, $[Sl(2)\times Sl(2)]/[SO(2)
\times SO(2)]\times R^+$ in 6, $[Sl(3)\times Sl(3)]/[SO(3)\times SO(3)]$ in 5 and $Sl(6)/SO(6)$ in 4. We
proceed using the same strategy of Ref. [16], that is we list all the 8-dimensional fields that can give rise
to forms after dimensional reduction. We use the same notation as in [16], labeling each field with numbers
denoting the number of antisymmetric spacetime indices in 8 dimensions. The index $\alpha$ is in the
fundamental of $Sl(2)$. We give here the final result, where the first column denotes the highest dimension
for which the corresponding fields give rise to forms after dimensional reduction:
$$\vbox{ \offinterlineskip\halign{
\strut#&\vrule#\quad& \hfil$#$\hfil& \quad\vrule#\quad&  \hfil#&\quad\vrule#\cr \noalign{\hrule} & &  D &
\quad fields & & \cr \noalign{\hrule}  & &  8 & \quad  $A_3^\alpha \quad A_6^{\alpha\beta}  $ & & \cr
\noalign{\hrule} & & 7 & \quad $h_1{}^1 \quad A_{5,1} \quad A_{7,1,1}^\alpha \quad A_{8,1}^\alpha \quad
A_{8,1}^{\alpha\beta\gamma}  $ & & \cr \noalign{\hrule} & & 6 & \quad $A_{7,2}^\alpha \quad A_{8,2,1,1} $ & &
\cr \noalign{\hrule} & & 5 & \quad $A_{6,3}^\alpha \quad  A_{7,3,2} \quad A_{8,3,1}(\times 2) \quad
A_{8,3,1}^{\alpha\beta}(\times 2) \quad A_{8,3,3,1}^\alpha
 $ & & \cr \noalign{\hrule} & & 4 & \quad $A_{7,4,1} \quad A_{7,4,1}^{\alpha \beta} \quad A_{7,4,4}^\alpha$ & & \cr
\noalign{\hrule} }}
$$
Observe that some of the fields in the list have multiplicity higher than one. All these fields, as well as
the corresponding multiplicities, were listed in Ref. [31].

Performing the dimensional reduction, one obtains the results that are summarised in the following table:
$$\vbox{ \offinterlineskip\halign{ \strut#&\vrule#\quad& \hfil$#$\hfil& \quad\vrule#\quad& \hfil$#$\hfil&
\quad\vrule#\quad& \hfil$#$\hfil&\quad\vrule#\quad& \hfil$#$\hfil&\quad\vrule#\quad&
\hfil$#$\hfil&\quad\vrule#\quad& \hfil$#$\hfil&\quad\vrule#\quad&   \hfil$#$\hfil&\quad\vrule#\quad&
\hfil$#$\hfil&\quad\vrule#\quad& \hfil$#$\hfil&\quad\vrule#\quad& \hfil#&\quad\vrule#\cr \noalign{\hrule} & &
\omit\hidewidth D \hidewidth& & \omit\hidewidth G \hidewidth& & \omit\hidewidth $A_1$ \hidewidth & &
\omit\hidewidth $A_2$ \hidewidth & & \omit\hidewidth $A_3$ \hidewidth  & &  \omit\hidewidth $A_4$ \hidewidth
& & \omit\hidewidth $A_5$ \hidewidth & & \omit\hidewidth $A_6$ \hidewidth & & \omit\hidewidth $A_7$
\hidewidth  & & \omit\hidewidth $A_8$ \hidewidth & \cr \noalign{\hrule} & & 8& & $$Sl(2)$$ & &  & &  & &
${\bf 2}$ & &  & &  & & ${\bf 3}$ & &  & &  \hfil & \cr \noalign{\hrule}& &  7 & & $$Sl(2)\otimes R^+$$  & &
{\bf 1} & & {\bf 2} & & {\bf 2} & & {\bf 1} & & {\bf 1} & & {\bf 2} & & {\bf 2} & &  \hfil & \cr & &  & &   &
&  & &  & &  & &  & & {\bf 3} & & {\bf 3} & & {\bf 2} & & \hfil & \cr & &  & &   & &  & &  & &  & &  & &  & &
& & {\bf 4} & & \hfil & \cr\noalign{\hrule} & &  6 & &
$${Sl(2)^2}\otimes R^+$$ & & {\bf 2x1} & & {\bf 2x2} & & {\bf 2x1} & & {\bf 1x1} & & {\bf 2x1} & & {\bf 1x3} & &  & &
\hfil & \cr & &   & &  & & {\bf 1x2} & &  & & {\bf 1x2} & & {\bf 3x1} & & {\bf 1x2} & & {\bf 2x4} & & & &
\hfil & \cr  & &   & &  & &  & &  & &  & & {\bf 1x3} & & {\bf 2x3} & & {\bf 2x2} & & & & \hfil & \cr  & &   &
&  & & & &  & &  & &  & & {\bf 3x2} & & {\bf 2x2} & & & & \hfil & \cr  & &   & &  & & & &  & &  & &  & &  & &
{\bf 2x2} & & & & \hfil & \cr & &   & &  & & & &  & &  & &  & &  & & {\bf 4x2} & & & & \hfil & \cr & &   & &
& & & &  & & & &  & &  & & {\bf 3x1} & & & & \hfil & \cr \noalign{\hrule}   & &  5 & &
$${Sl(3)}\otimes Sl(3)$$ & & {\bf \overline{3}x\overline{3}} & & {\bf 3x3} & & {\bf 8x1} & & {\bf \overline{3}x6}
& & \omit\hidewidth {\bf 15x3} \hidewidth & &  & &  & & \hfil & \cr & &   & &  & &  & &  & & {\bf 1x8} & &
{\bf 6x\overline{3}} & & \omit\hidewidth  ${\bf 3x{15}}$ \hidewidth & &  & & & & \hfil & \cr  & &   & &  & &
& & & & & & {\bf \overline{3}x\overline{3}} & & {\bf \overline{6}x3} & &  & & & & \hfil & \cr  & &   & &  & &
& & & & & &  & & {\bf 3x3} & & & & & & \hfil & \cr  & &   & &  & & & & & & & &  & & {\bf 3x3} & & & & & &
\hfil & \cr & &   & & & & & & & &  & & & & {\bf 3 x \overline{6}} & & & & & & \hfil & \cr \noalign{\hrule} &
&  4 & &
$$Sl(6)$$ & & {\bf 20} & & {\bf 35} & & {\bf 70} & & ?
& &  & &  & &  & & \hfil & \cr & &   & &  & &  & &  & & {\bf \overline{70}} & &  & &  & &  & & & & \hfil &
\cr \noalign{\hrule}}}
$$

The $D-1$-forms in the table correspond to the massive deformations that the non-linear realisation allows in
dimension $D$.

\vskip 1cm \noindent {\bf B \ \ $F_4^{+++}$ in lower dimensions}
\medskip
\noindent In this Appendix we perform for the supersymmetric $F_4^{+++}$ case the same analysis that was
carried out in the previous Appendix for the $E_6^{+++}$ case. We determine all the forms that arise in
dimensions from 6 to 4, with the exception of the 4-forms in 4 dimensions.

In six dimensions, the theory describes the bosonic sector of the gravity multiplet together with two tensor
multiplets and two vector multiplets. There are two scalars, belonging to the tensor multiplets,
parametrising the manifold $Sl(2)/SO(2)$. In five dimensions this corresponds to the gravity multiplet plus
five vector multiplets, and the 5 scalars parametrise $Sl(3)/SO(3)$. Finally, in four dimensions the theory
describes an ${\cal N}=2$ gravity multiplet together with 6 vector multiplets, and the 12 scalars parametrise
the manifold $Sp(6)/[SU(3)\otimes U(1)]$.

The list of all the fields that give rise to forms after dimensional reduction, together with the highest
dimension for which this occurs, is given here:
$$\vbox{ \offinterlineskip\halign{
\strut#&\vrule#\quad& \hfil$#$\hfil& \quad\vrule#\quad&  \hfil#&\quad\vrule#\cr \noalign{\hrule} & &  D &
\quad fields & & \cr \noalign{\hrule}  & &  6 & \quad  $A_1^\alpha \quad A_2^{\alpha\beta} \quad A_3^\alpha
\quad A_4^{\alpha\beta} \quad A_5^\alpha \quad A_5^{\alpha\beta\gamma} \quad A_6^{\alpha\beta}(\times 2)
\quad A_6^{\alpha \beta \gamma \delta}$ & & \cr \noalign{\hrule} & &  5 & \quad   $ h_1{}^1 \quad A_{3,1}
\quad A_{4,1}^\alpha \quad A_{5,1}(\times 2) \quad A_{5,1}^{\alpha\beta} \quad A_{6,1}^\alpha (\times 3 )
\quad A_{6,1}^{\alpha\beta\gamma}(\times 2 ) $ & & \cr & & & $\quad A_{5,1,1}^\alpha \quad
A_{6,1,1}^{\alpha\beta}(\times 2) \quad A_{6,1,1}(\times 2) \quad A_{6,1,1,1}^\alpha $ & & \cr
\noalign{\hrule} & & 4 & \quad $A_{4,2}^{\alpha\beta} \quad A_{5,2}^\alpha (\times 2) \quad
A_{5,2}^{\alpha\beta\gamma} \quad A_{5,2,1} \quad A_{5,2,1}^{\alpha\beta} \quad A_{5,2,2}^\alpha $ & & \cr
\noalign{\hrule}}}
$$

Observe that some of the fields in the list have multiplicity higher than one. Although most of the fields
where listed in the tables of [9,31] up to multiplicity 6, the fields with more that 6 indices have not
appeared in the literature.

By dimensional reduction, one can then obtain all the forms that arise. The results are summarised in the
Table:
$$\vbox{ \offinterlineskip\halign{
\strut#&\vrule#\quad& \hfil$#$\hfil& \quad\vrule#\quad& \hfil$#$\hfil& \quad\vrule#\quad&
\hfil$#$\hfil&\quad\vrule#\quad& \hfil$#$\hfil&\quad\vrule#\quad&   \hfil$#$\hfil&\quad\vrule#\quad&
\hfil$#$\hfil&\quad\vrule#\quad&   \hfil$#$\hfil&\quad\vrule#\quad& \hfil#&\quad\vrule#\cr \noalign{\hrule} &
& \omit\hidewidth D \hidewidth& & \omit\hidewidth G \hidewidth& & \omit\hidewidth $A_1$ \hidewidth & &
\omit\hidewidth $A_2$ \hidewidth & & \omit\hidewidth $A_3$ \hidewidth  & &  \omit\hidewidth $A_4$ \hidewidth
& & \omit\hidewidth $A_5$ \hidewidth & & \omit\hidewidth $A_6$ \hidewidth &  \cr \noalign{\hrule} & & 6& &
$$Sl(2)$$ & & {\bf 2} & & {\bf 3} & & ${\bf 2}$ & & {\bf 3} & & {\bf 2} & & ${\bf 3}$  \hfil & \cr  & &  & &
& &  & &  & &  & &  & & {\bf 4} & & ${\bf 3}$  \hfil & \cr  & &  & &  & &  & &  & &  & &  & &  & & ${\bf 5}$
\hfil & \cr \noalign{\hrule}& &  5 & &
$$Sl(3)$$ & & {\bf \overline{6}} & & {\bf 6} & & {\bf 8} & & {\bf 3} & & {\bf \overline{15}} & &    \hfil & \cr
& & & &   & & & & & & & & {\bf 15} & & {\bf \overline{24}} & &   \hfil &\cr & & & &   & & & & & & & & & &
{\bf 6} & &   \hfil &\cr & & & &   & & & & & & & &  & & {\bf \overline{3}} & & \hfil & \cr \noalign{\hrule} &
&  4 & &
$$Sp(6)$$ & & {\bf 14'} & & {\bf 21} & & {\bf 64} & & ? & &  & &    \hfil & \cr  \noalign{\hrule} }}
$$
The table shows in particular that the six-dimensional theory possesses massive deformations in the ${\bf 2
\oplus 4}$ of $SL(2)$, the five-dimensional one in the ${\bf \overline{3} \oplus \overline{15}}$ of $Sl(3)$,
and the four-dimensional one in the ${\bf 64}$ of $Sp(6)$.

In the original version of this paper, the forms listed in the last table of this appendix for $F_4^{+++}$
and those listed in the last table of appendix A for $E_6^{+++}$, as well as the forms obtained in section 6
for $G_2^{+++}$, were computed using the tables of Refs. [9] and [31]. However, it turns out that for some of
the higher rank forms these tables are not sufficiently exhaustive to contain all the required fields. We
have used the program SimpLie [19], available on http://strings.fmns.rug.nl/SimpLie/, to compute the extra
fields and have subsequently modified the tables and corrected section 6. We thank the referee for drawing
this to our attention.

\vskip 1cm \noindent {\bf C \ \ $G_2^{+++}$ and minimal gauged supergravity in five dimensions}
\medskip
\noindent The minimal supergravity multiplet in five dimensions contains the metric, a $U(1)$ vector $A_\mu$
and a $USp(2)$ doublet of gravitini $\psi_\mu^i$ satisfying symplectic Majorana conditions. The $USp(2)$
symmetry only acts on the gravitini, and in [32] it was shown that this theory admits a gauging of a $U(1)$
subgroup of the $USp(2)$ symmetry. This gauging is thus different from the ones that occur in maximal
supergravity because it arises from a symmetry that only acts on the fermions.

The field equation for the vector at lowest order in the fermions, in both the massless and the gauged case,
has the form
  $$
  D_\mu F^\mu{}_\nu = - {1 \over 2 \sqrt{6}}\sqrt{-g} \epsilon_{\mu \mu_1 \dots\mu_4} F^{\mu_1 \mu_2}
F^{\mu_3\mu_4} \eqno(C.1)
  $$
because of the presence of a Chern-Simons term $A\wedge F \wedge F$ in the lagrangian, where $F_{\mu\nu}$ is
the field strength of $A_\mu$. In the gauged theory the lagrangian contains a minimal coupling $g$ of the
gravitino to the vector, as well as a mass term $g$ for the gravitino and a cosmological constant $-g^2$. The
lagrangian reads, up to quartic order fermi terms,
  $$
  {\cal L} = \sqrt{-g} [-{1 \over 2} R - {1 \over 4} F_{\mu\nu} F^{\mu\nu} -{1 \over 2}
  \bar{\psi}^i_\mu \gamma^{\mu\nu\rho}
  D_\nu \psi_{\rho i}
  +{1 \over 2} g \bar{\psi}^i_\mu \gamma^{\mu\nu\rho} A_\nu \delta_{ij} \psi_{\rho}^j  -i{\sqrt{6} \over 4} g
\bar{\psi}^i_\mu
  \gamma^{\mu\nu} \psi_\nu^j \delta_{ij}
  $$
  $$
   + 4 g^2- {3 \over 8 \sqrt{6}}i \bar{\psi}_\mu^i (\gamma^{\mu\nu\rho\sigma} + 2 g^{\mu\nu} g^{\rho\sigma})
\psi_{\sigma i} F_{\nu \rho}]
  + {1 \over 6\sqrt{6}}\epsilon^{\mu\nu\rho \sigma\tau} A_\mu F_{\nu\rho}  F_{\sigma \tau} \quad .
\eqno(C.2)
  $$
The fermionic terms that arise in the gauged theory, that is the order $g$ terms in eq. (C.2), are
proportional to $\delta_{ij}$, which breaks $USp(2)$ explicitly.

The authors of [35] extended the algebra of this model introducing a 2-form dual to the vector, as well as
3-forms and 4-forms that are both non-propagating. In five dimensions  a 3-form is dual to a scalars but in
[35]  the 3-forms are introduced regardless the fact that there are no scalars in the model, and therefore
their field strength is required to vanish identically. The authors originally showed that the supersymmetry
algebra allows the introduction of a 3-form and a 4-form whose field strength is dual to the coupling
constant $g$ in the gauged theory. In the first version of this paper, we pointed out that this would have
led to an explicit symmetry breaking of $USp(2)$ in the algebra in the ungauged theory, and an explicit
breaking of the $R$ symmetry is inconsistent with supersymmetry. The authors then revised their paper showing
that both the 3-form and the 4-form are actually contained in a $USp(2)$ triplet, and inserting a footnote to
acknowledge our contribution.

The $G_2^{+++}$ non-linear realisation in five dimensions  describes the bosonic sector of this model, and in
particular it contains a vector and its dual 2-form [9]. No 3-forms and 4-forms are present, and the authors
of [35] pointed out that the mass parameter of the gauged supergravity of [32] can not be obtained as the
dual of a 4-form arising in the $G_2^{+++}$ non-linear realisation, because no such forms are present in this
theory in 5 dimensions.

We now summarise the results of ref. [35]. In the ungauged theory,  the supersymmetry algebra closes on the
2-form $B_{\mu\nu}$ as expected from $G_2^{+++}$, imposing that its field strength
  $$
  G_{\mu\nu\rho} = 3\partial_{[\mu} B_{\nu\rho]} -\sqrt{6} A_{[\mu} F_{\nu\rho]}
\eqno(C.3)
  $$
is dual to the field strength of the vector $F_{\mu\nu}$. Taking the curl of the duality relation
  $$
  G_{\mu\nu\rho} = {1 \over 2} \epsilon_{\mu\nu\rho\sigma\tau} F^{\sigma \tau} \eqno(C.4)
  $$
one recovers the vector field equation (C.1). The $USp(2)$ triplet of 3-forms $C_{\mu\nu\rho}^{ij}$ not found
in $G_2^{+++}$ transforms under supersymmetry as
  $$
  \delta C^{ij}_{\mu\nu\rho} = i\bar{\epsilon}^{(i} \gamma_{[\mu \nu} \psi_{\rho
]}^{j)}\eqno(C.5)
  $$
and the commutator of two supersymmetry transformations closes at lowest order in the fermions imposing that
the field strength vanishes identically,
  $$
  H_{\mu\nu\rho\sigma}^{ij}  = \partial_{[\mu} C_{\nu\rho\sigma]}^{ij} =0 \quad .
\eqno(C.6)
  $$
The 4-forms $D^{ij}_{\mu\nu\rho\sigma}$ transform under supersymmetry as
$$
  \delta D^{ij}_{\mu\nu\rho\sigma} = \bar{\epsilon}^{(i} \gamma_{[\mu \nu\rho}
\psi_{\sigma ]}^{j)}
   - i \sqrt{6}A_{[\mu} \bar{\epsilon}^{(i} \gamma_{\nu\rho} \psi_{\sigma ]}^{j)} \eqno(C.7)
  $$
and their field strength vanishes identically, {\it i.e.}
$$
  L_{\mu\nu\rho\sigma\tau}^{ij}  = \partial_{[\mu} D_{\nu\rho\sigma \tau]}^{ij} =0 \quad
. \eqno(C.8)
  $$
The supersymmetry transformation of the gravitino has the form [32]
  $$
  \delta \psi_\mu^i = D_\mu \epsilon^i + i {1 \over 4 \sqrt{6}} ( \gamma_\mu{}^{\nu\rho} - 4 \delta_\mu^\nu
\gamma^\rho )F_{\nu\rho}
  \epsilon^i \eqno(C.9)
  $$
and at lowest order in the fermions one has to only consider the variations of the gravitino in eqs. (C.5)
and (C.7). This only produces gauge transformations, while the general coordinate transformations are
generated from the fact that the field strengths of eq. (C.6) and (C.8) vanish identically. Thus, the fact
that the field strengths vanish identically implies that these fields are pure gauge, and thus supersymmetry
closes on these fields in a rather trivial way, as the authors point out [35].

The procedures used in [35] and the ones  used in the $E_{11}$ formulation of maximal supergravities are
different for a number of reasons. First of all, in Ref. [35] the introduction of 3-forms and 4-forms is
trivial in the sense that their field strengths of the 3-form and the 4-form do not contain any lower rank
field. This is in contrast with the $E_{11}$ case, in which the field strengths of higher rank contain the
fields of lower rank. Secondly, while in $E_{11}$ the $D-2$ forms are dual to scalars, and their field
strengths are not identically zero, in the model of [35] there is no real democracy because there are no
scalars and this makes the introduction of 3-forms somehow artificial. Finally, the 3-forms and 4-forms are
triplets of the fermionic global symmetry $USp(2)$ and thus are not related to the bosonic symmetry that
arises in the non-linear realisation (which is absent in this five-dimensional case). The closure of the
supersymmetry algebra does rely on properties of the $\gamma$ matrices. However, the 3-forms and 4-forms do
not couple to the other fields and thus the closure is a rather trivial consequence of the $\gamma$ matrix
identities. The democratic formulation of the supersymmetry algebra of maximal five dimensional supergravity
theory has recently been described in [37]. In that case the 3-forms are in the adjoint of $E_6$, and are
dual to the scalars that realise non-linearly $E_6$ with local subalgebra $USp(8)$. One can see from that
result that one needs the same $\gamma$ matrix identities to cancel the $F_{\mu\nu}$ terms in the
supersymmetry commutator on the 3-forms, and thus putting to zero the scalars and the field strengths of the
3-forms is from this point of view a singular limit, in which 3-forms arise because one is using only a part
of the constraints that come from the algebra when the scalars are present.

We now study the gauged theory of [35]. The supersymmetry variation of the gravitino is modified by the
addition of a term of the form [32]
  $$
  \delta^\prime \psi_\mu^i = - g A_\mu \delta^{ij} \epsilon_j - i{1 \over \sqrt{6}} g \gamma_\mu \delta^{ij}
\epsilon_j \quad .
  \eqno(C.10)
  $$
This term does not affect the supersymmetry commutator on the 3-forms, while it does affect the commutator on
the 4-forms, which now closes provided that the duality relation
  $$
  L_{\mu\nu\rho\sigma\tau}^{ij} \sim g \epsilon_{\mu\nu\rho\sigma \tau}
\delta^{ij}\eqno(C.11)
  $$
holds, where $L_{\mu\nu\rho\sigma\tau}^{ij}$ is the gauge invariant field strength defined in (C.8). The
supersymmetry algebra on the 2-form implies that the gauge invariant field strength for the 2-form is now
  $$
  G_{\mu\nu\rho} = 3\partial_{[\mu} B_{\nu\rho]} -\sqrt{6} A_{[\mu} F_{\nu\rho]} -3 g \delta_{ij}
C^{ij}_{\mu\nu\rho}
  \quad .
  \eqno(C.12)
  $$
Thus, the 2-form transforms with respect to the gauge parameter of the 3-form, which therefore can be used to
gauge away the 2-form completely. Secondly, the field strengths of the 4-forms are dual to the coupling
constant $g$. We observe that the field strengths of the 3-forms and the 4-forms are not modified with
respect to the massless case, and in particular the 3-forms are pure gauge quantities also in the gauged
case. Again, like in the massless case there is no real democracy, and in particular one can introduce the
4-forms dual to the mass parameters without needing to introduce the 2-form and the 3-forms. This is in
contrast with what happens in the $E_{11}$ non-linear realisation. Finally, if one believes that $D-1$ forms
are responsible for gauged supergravities, the existence of a triplet of 4-forms as proposed in [35] would
predict three such theories. However, only one such theory exists. This is again in contrast with what
happens in the $E_{11}$ case, in which the number of deformations and the number of $D-1$ forms coincide in
any dimension.

We believe that there are exceptional cases for which the gauged theory is not accounted for by the bosonic
degrees of freedom in $G^{+++}$ but by the fermionic partners as treated from the view point of the
non-linear realisation. This particular model is singular in the sense that it has no scalars and has a
global symmetry that only acts on the fermions.

As a toy model showing that the analysis of higher rank forms becomes more singular in cases with fewer
supersymmetry and no scalars, we consider the very well known theory of pure minimal supergravity in 4
dimensions. In this case the supergravity multiplet only contains the metric and a Majorana gravitino, and
one can introduce a negative cosmological constant $-g^2$ and a mass term for the gravitino, that
schematically correspond to the appearance in the action of the terms
  $$
  {\rm det}e (g \bar{\psi}_\mu \gamma^{\mu\nu} \psi_\nu + g^2 ) \eqno(C.13)
  $$
and supersymmetry is provided by the fact that the gravitino transforms as
  $$
  \delta \psi_\mu = D_\mu \epsilon + g \gamma_\mu \epsilon \quad .
  \eqno(C.14)
  $$
We first want to consider the introduction of 2-forms. These would be the equivalent of 3-forms in five
dimensions. In the massless theory the supersymmetry algebra closes trivially on a 2-form $B_{\mu\nu}$ whose
supersymmetry transformation is
  $$
  \delta B_{\mu\nu} = \bar{\epsilon} \gamma_{[\mu} \psi_{\nu ]}\eqno(C.15)
  $$
if one imposes
  $$
  \partial_{[\mu} B_{\nu\rho ]} =0 \quad . \eqno(C.16)
  $$
In the massive theory this field can no longer be introduced because the supersymmetry commutator produces a
term of the form $ g \bar{\epsilon}_1 \gamma_{\mu\nu} \epsilon_2$ which can not be interpreted as a gauge
transformation of any kind.

We now move to the 3-forms.  One can introduce in both the massless and the massive theory a 3-form
$C_{\mu\nu\rho}$ whose supersymmetry variation is
  $$
  \delta C_{\mu\nu\rho} = \bar{\epsilon} \gamma_{[\mu\nu} \gamma_5 \psi_{\rho
]}\eqno(C.17)
  $$
if one imposes
  $$
  \partial_{[\mu} C_{\nu\rho \sigma]} =  g \epsilon_{\mu\nu\rho\sigma}  \quad .
\eqno(C.18)
  $$
This form is therefore dual to the mass parameter $g$. In addition to this, one can also introduce a 3-form
$C^\prime_{\mu\nu\rho}$ whose supersymmetry variation is
  $$
  \delta C^\prime_{\mu\nu\rho} = \bar{\epsilon} \gamma_{[\mu\nu} \psi_{\rho ]}\eqno(C.19)
  $$
if one imposes
  $$
  \partial_{[\mu} C^\prime_{\nu\rho \sigma]} = 0 \quad . \eqno(C.20)
  $$
This additional 3-form has therefore vanishing field strength in both the massless and the massive theory.
Therefore, in this model one can introduce a 2-form only if $g=0$, while for any value of the coupling
constant one can introduce two 3-forms, one of them being dual to $g$. While it seems that for any massive
supersymmetric theory one can introduce a $D-1$ form  whose field strength is dual to the mass deformation
parameter, we think that this model  reveals the singular nature of these manipulations in theories that are
not enough constrained by supersymmetry.

\vskip 1cm \noindent{\bf References}
\medskip
\item{[1]}
 I.~C.~G.~Campbell and P.~C.~West,
 {\it N=2 D = 10 Nonchiral Supergravity  And Its Spontaneous
Compactification},
 Nucl.\ Phys.\ B {\bf 243} (1984) 112;
 F.~Giani and M.~Pernici,
 {\it N=2 Supergravity In Ten-Dimensions},
 Phys.\ Rev.\ D {\bf 30} (1984) 325;
 M.~Huq and M.~A.~Namazie,
 {\it Kaluza-Klein Supergravity In Ten-Dimensions},
 Class.\ Quant.\ Grav.\  {\bf 2} (1985) 293
 [Erratum-ibid.\  {\bf 2} (1985) 597].

\item{[2]} J. Schwarz and  P. West {\it Symmetries and   Transformations of
   chiral N=2, D=10 super\-gravity}, Phys. Lett. {\bf B126}  (1983),  301.

\item{[3]}  P. Howe and  P. West,    {\it The complete N=2, d=10
supergravity}, Nucl. Phys. {\bf B238}
   (1984) 181.

\item{[4]}  J. Schwarz, {\it Covariant field equations of chiral N=2   D=10
   supergravity}, Nucl. Phys. {\bf B226} (1983), 269.

\item{[5]}
 E.~Cremmer, B.~Julia and J.~Scherk,
 {\it Supergravity theory in 11 dimensions},
 Phys.\ Lett.\  B {\bf 76} (1978) 409.

\item{[6]} P.~C. West, {\it Hidden superconformal symmetry in {M}
  theory },  JHEP {\bf 08} (2000) 007, {\tt hep-th/0005270}

\item{[7]} P. West, {\it $E_{11}$ and M Theory}, Class. Quant.
Grav. {\bf 18 } (2001) 4443,  hep-th/010408.

\item{[8]} I. Schnakenburg and P. West, {\it Kac-Moody Symmetries of
IIB supergravity}, Phys. Lett. {\bf B 517} (2001) 137-145, {\tt hep-th/0107181}

\item{[9]} A. Kleinschmidt, I. Schnakenburg and P. West, {\it
Very-extended Kac-Moody algebras and their interpretation at low  levels}, {\tt hep-th/0309198}

\item{[10]} L.~J.~Romans, {\it Massive N=2a Supergravity In
Ten-Dimensions},  Phys.\ Lett.\  B {\bf 169} (1986) 374.

\item{[11]} P. West, {\it  The  IIA, IIB and eleven dimensional
theories and their common $E_{11}$ origin}, hep-th/0402140.

\item{[12]} E. Bergshoeff, Mess de Roo, S. Kerstan and F. Riccioni, {\it
IIB Supergravity Revisited}, hep-th/0506013.

\item{[13]} P. West, {\it  $E_{11}$, ten forms and supergravity }, JHEP
0603 (2006) 072, hep-th/0511153.

\item{[14]} Eric A. Bergshoeff, Mees de Roo, Sven F.
Kerstan, T. Ortin,  Fabio Riccioni, {\it  IIA Ten-forms and the Gauge  Algebras of Maximal Supergravity
Theories},   JHEP 0607 (2006) 018, hep-th/0602280.

\item {[15]}   F.~Riccioni and P.~West, {\it Dual fields and E(11)},
 Phys.\ Lett.\  B {\bf 645} (2007) 286, arXiv:hep-th/0612001.

\item {[16]} F.  Riccioni and P. West, {\it The $E_{11}$ origin of all
maximal supergravities}, arXiv:0705.0752.

\item{[17]}  B.~de Wit, H.~Samtleben and M.~Trigiante,
 {\it On Lagrangians and gaugings of maximal supergravities},
 Nucl.\ Phys.\  B {\bf 655} (2003) 93
 [arXiv:hep-th/0212239].

\item{[18]}  A. Sagnotti, {\it Open strings and their symmetry groups},
hep-th/0208020.

\item{[19]} E. Bergshoeff, I. De Baetselier and T. Nutma, {\it $E_{11}$
and the Embedding Tensor}, ArXiv :0705.130.

\item{[20]} N.D. Lambert, P.C. West, {\it Coset Symmetries in
Dimensionally Reduced Bosonic String Theory},  Nucl.Phys. B615 (2001)  117-132, hep-th/0107209.

\item{[21]} I. Schnakenburg and P. West, {\it Kac-Moody Symmetries  of  the
Ten-dimensional  Non-maximal Supergravity Theories}, JHEP 0405  (2004)  019, hep-th/0401196.

\item{[22]} M. R. Gaberdiel, D. I. Olive and P. West, {\it A class of
Lorentzian Kac-Moody algebras}, Nucl. Phys. {\bf B 645} (2002)  403-437, hep-th/0205068.

\item{[23]} F. Englert, L. Houart, A. Taormina, P. West, {\it  The
Symmetry of M-Theories},  JHEP 0309 (2003) 020, hep-th/0304206.

\item{[24]} F. Englert, L. Houart, P. West, {\it Intersection rules,
dynamics and symmetries},  JHEP 0308 (2003) 025, hep-th/0307024.

\item{[25]}  T. Damour, M. Henneaux and H. Nicolai, {\it $E_{10}$ and a
``small tension expansion'' of M-theory}, Phys. Rev. Lett. {\bf  89} (2002) 221601, hep-th/0207267.

\item{[26]} P. West, {\it Very Extended $E_8$ and $A_8$ at low levels,
Gravity and Supergravity}, Class. Quant. Grav. 20 (2003) 2393-2406,  hep-th/0212291.

\item{[27]} V.~G.~Kac,
 {\it Infinite dimensional Lie algebras},  Cambridge, UK: Univ. Pr.
(1990).

\item {[28]}  E. Cremmer, B. Julia, H. Lu and  C.N. Pope, {\it
Higher-dimensional Origin of D=3 Coset Symmetries}, arXiv:hep-th/ 9909099.

\item{[29]} E.~Cremmer, {\it Supergravities In 5 Dimensions},
in {\it Superspace and Supergravity}, Eds. S.W. Hawking and M. Rocek   (Cambridge Univ. Press, 1981).

\item{[30]}
 A.~H.~Chamseddine and H.~Nicolai,
 {\it Coupling The SO(2) Supergravity Through Dimensional Reduction},
 Phys.\ Lett.\  B {\bf 96} (1980) 89.

\item{[31]} A. Kleinschmidt and P. West, {\it  Representations of G+++ and
the role of space-time},  JHEP 0402 (2004) 033,  hep-th/0312247.

\item{[32]} M. Gunaydin, G. Sierra and P. Townsend, Nucl. Phys. B 253,
(1985) 573.

\item{[33]}
S.~Mizoguchi and N.~Ohta,
 {\it More on the similarity between D = 5 simple supergravity and M theory},
 Phys.\ Lett.\  B {\bf 441} (1998) 123
 [arXiv:hep-th/9807111].

\item{[34]}  F.~Riccioni and P.~West, to be published.

\item{[35]} J. Gomis and D. Roest, {\it Non-propagating Degrees of Freedom
in supergravity and very extended $G_2$}, arXiv:0706.0667.

\item{[36]}
  A.~Kleinschmidt and H.~Nicolai,
  {\it IIB supergravity and E(10)},
  Phys.\ Lett.\  B {\bf 606} (2005) 391
  [arXiv:hep-th/0411225].

\item{[37]}
  F.~Riccioni and P.~West,
  {\it E(11)-extended spacetime and gauged supergravities},
  arXiv: 0712.1795 [hep-th].

\end